\documentclass[a4paper,12pt]{article}
\usepackage{amssymb}
\usepackage[usenames,dvipsnames]{color}
\usepackage{slashed}
\usepackage[table]{xcolor}
\usepackage{graphicx}
\usepackage{mathtools}
\usepackage{cite} 
\usepackage{hyperref}
\usepackage{comment}
\usepackage{enumerate}
\usepackage[top=30mm, bottom=30mm, right=25mm, left=25mm, headheight=15.72pt]{geometry}

\usepackage{amsthm}
\usepackage{bm}

\theoremstyle{plain}

\theoremstyle{definition}

\theoremstyle{remark}

\usepackage{chngcntr}

\counterwithin*{equation}{section}

\def\cL{{\cal L}}
\def\cT{{\cal T}}

\title{\bf{Fractons in effective field theories for spontaneously broken translations}}
\date{}
\author{Riccardo Argurio$^{a,}$\footnote{riccardo.argurio@ulb.be}, 
Carlos Hoyos$^{b,}$\footnote{hoyoscarlos@uniovi.es}, 
Daniele Musso$^{c,}$\footnote{daniele.musso@usc.es}~ and 
Daniel Naegels$^{a,}$\footnote{daniel.naegels@ulb.be}}

\begin{document}
\maketitle
\begin{center}\it{
$^{a}$Physique Th\'eorique et Math\'ematique and International Solvay Institutes, \\ Universit\'e Libre de Bruxelles, C.P. 231, B-1050 Brussels, Belgium\\
\vspace{15pt}
$^{b}$Department of Physics and\\ 
Instituto de Ciencias y Tecnolog\'ias Espaciales de Asturias (ICTEA)\\
 Universidad de Oviedo,  c/ Federico Garc\'{\i}a Lorca 18, E-33007 Oviedo, Spain\\
\vspace{15pt}
$^{c}$Centro de Supercomputaci\'on de Galicia (CESGA),\\
s/n, Avenida de Vigo, 15705 , Santiago de Compostela, Spain\\
}
\end{center}
\vspace{25pt}

\begin{abstract}
\noindent
We study the concomitant breaking of spatial translations and dilatations in Ginzburg-Landau-like models, where the dynamics responsible for the symmetry breaking is described by an effective Mexican hat potential for spatial gradients. We show that there are fractonic modes with either subdimensional propagation or no propagation altogether, namely, immobility.
Such class of effective field theories encompasses instances of helical superfluids and meta-fluids, where fractons can be connected to an emergent symmetry under higher moment charges, leading in turns to the trivialization of some elastic coefficients.
The introduction of a finite charge density alters the mobility properties of fractons and leads to a competition between the chemical potential and the superfluid velocity in determining the gap of the dilaton.
The mobility of fractons can also be altered at zero density upon considering additional higher-derivative terms.

\end{abstract}
\newpage

\tableofcontents

\section{Introduction}

An interesting aspect of low-energy effective theories is that of emergent symmetries. In the simplest setup of a complex scalar field with a Mexican hat potential, the $U(1)$ symmetry associated to phase rotations of the scalar is spontaneously broken and the low-energy effective theory is described by a massless Nambu-Goldstone boson. At sufficiently low energies, the effective action of the theory is that of a massless scalar field, which not only enjoys the original $U(1)$ symmetry in the form of a constant shift of the Nambu-Goldstone field, but it is also conformal invariant and has an infinite set of conserved higher-spin currents associated to coordinate-dependent shifts of the Nambu-Goldstone field. Neither the conformal nor the coordinate-dependent shifts are symmetries of the full theory, and they are broken when higher-derivative corrections to the low-energy action are considered. Nevertheless, they can leave an imprint in the properties of the low-energy effective theory. 

Similar emergent symmetries at low energies appear in other contexts like low-energy excitations of a Fermi surface, independent spin and spatial rotation symmetries in non-relativistic theories, etc. Here we want to explore low-energy effective theories with emergent symmetries that lead to (gapless) fractonic modes. Fractons are excitations that are able to move only along a restricted set of spatial directions, or are even completely immobile \cite{Pretko:2020cko,nandkishore_fractons_2019}. Gapless fractons appear in a variety of systems such as spin liquids \cite{Xu2006,Xu2010,Pretko2017spinliquid,Pretko2017spinliq2,You_emergent_2020},
dipole-conserving lattice models
\cite{Pai2019,Feldmeier2020,Morningstar2020,Iaconis2021,moudgalya2021spectral} and
quantum elasticity
\cite{PretkoSolid2018,Gromov2019elastic,Kumar2019,Pretko:2018fed,pretko2019crystal,zhai2019two,Gromov:2019waa,Nguyen:2020yve,Fruchart2020dual,Manoj:2020abe,Surowka:2021ved,Kleman2008review}. Hydrodynamics of fractons has been studied in \cite{Aasen:2020zru,grosvenor2021hydrodynamics,Glorioso:2021bif}. Models with spontaneous breaking of symmetries have also been studied \cite{Yuan:2019geh,Chen2021}. At low energies, the models we are going to discuss have similarities to these last, but with the important difference that it is not necessary to impose any exact coordinate-dependent phase rotation or shift symmetry in order to obtain fractonic dispersion relations.

A second aspect that we want to explore is the effect of spontaneous breaking of spacetime symmetries in the counting of Nambu-Goldstone bosons. It is well known that the naïve counting of a number of gapless modes equal to the number of broken generators does not apply in this case \cite{Ivanov:1975zq,Low:2001bw,Watanabe:2013iia,Brauner:2014aha}. An interesting case is when time translations are broken by a finite chemical potential. Under these conditions some of the Nambu-Goldstone bosons become gapped when the effective unbroken Hamiltonian does not commute with some of the broken generators  \cite{Miransky:2001tw,Schafer:2001bq,Nicolis_2013,Nicolis:2013sga,Watanabe:2013uya,Endlich_2015,Argurio:2020jcq}. In particular, if scale invariance is spontaneously broken together with a global symmetry, the dilaton will get a gap proportional to the chemical potential \cite{Argurio:2020jcq} since the generator of dilatations does not commute with the Hamiltonian. Integrating out the gapped modes and keeping only the gapless modes would be equivalent to applying the inverse Higgs constraints \cite{Ivanov:1975zq,Brauner:2014aha}.

If, instead of time translations, space translations are broken, we expect to find some qualitative similarities. There will be unbroken generators of space translations of the form $\widetilde{P}_i=P_i-k_{ia} Q^a$, where $P_i$ are the ordinary generators of space translations and $Q^a$ are the generators of spontaneously broken global symmetries. The generator of dilatations $D$ does not commute with the unbroken generators $[D,\widetilde{P}_i]=i P_i$, so this might produce a gap for the dilaton dependent on $k_{ia}$. However, due to the breaking of spatial symmetries, the dispersion relations of the modes can depend in a non-trivial way on the spatial momenta, so the intuition from the chemical potential does not entirely apply to this more complicated situation. 

What we will do in this work is to examine these questions using a simple $2+1$-dimensional model which can be viewed as a generalization of the ordinary Mexican hat model for spatial derivative terms of a complex scalar field. Scale invariance is ensured by introducing an additional real scalar acting as a compensating field. It turns out that there is a large space of possible ground states breaking translation invariance, and the effective theory depends crucially on the symmetry realization of the ground state. We restrict to states leading to homogeneous effective theories.
We find emergent symmetries leading to fractonic dispersion relations and a strong dependence on spatial momentum that affects both the dispersion relations and the composition of the modes. We also study generalizations to finite chemical potential and to $3+1$ dimensions for some cases.

The paper is organized as follows. In Section~\ref{sec:model} we introduce the model and discuss its ground states and symmetries. In Section~\ref{sec:dispersion} we compute the dispersion relations for linearized fluctuations around the ground states and identify the associated Nambu-Goldstone modes. In Section~\ref{OrignModelFiniteDensity} we extend our results to finite density and in Section~\ref{sec:G} we introduce a deformation that removes some of the degeneracy of the simpler model and study its effect on the dispersion relations. In Section~\ref{sec:count} we try to compare our results with theorems determining the number of gapless Nambu-Goldstone bosons and finally we conclude in Section~\ref{sec:discuss} with a discussion of possible physical systems where similar symmetry realizations and exotic Nambu-Goldstone modes might be found. We have collected several technical results and generalizations to $3+1$ dimensions in the Appendices.

\section{Translation-breaking Mexican hat model}\label{sec:model}

We consider a (2+1)-dimensional model with two scalar fields, one complex and one real, governed by the following Lagrangian density
\begin{equation}\label{lagra}
 {\cal L} =
 \partial_t \Phi^* \partial_t \Phi
 + A \partial_i \Phi^* \partial_i \Phi
 +\frac{1}{2} \partial_t \Xi \partial_t \Xi
 -\frac{1}{2} \partial_i \Xi \partial_i \Xi
 -B \frac{\left(\partial_i \Phi^* \partial_i \Phi\right)^2}{\Xi^6}
- H \Xi^6 \ .
\end{equation}
The ``couplings" $A$, $B$ and $H$ are all real and positive. The real scalar field $\Xi$ presents a standard kinetic term and plays the role of a ``compensator field," introduced in order to ensure scale invariance. The detailed scaling dimensions of the couplings and of the fields (and of the expectation values that we will introduce below) are
\begin{equation}
 [A] = [B] = [H] = 0\ , \qquad
 [\Phi] = [\rho] = \frac{1}{2}\ , \qquad 
 [\Xi] = [v] = \frac{1}{2}\ , \qquad 
 [k] = [\partial] = 1\ ,
\end{equation}
where we considered natural units of energy.

The complex scalar field $\Phi$ presents instead a non-standard kinetic term. Specifically, given the positivity of $A$, the quadratic term with spatial gradients has the opposite sign with respect to the standard relativistic action. This is a key ingredient for triggering the breaking of translation symmetry through configurations with non-vanishing gradients. Intuitively, the ``wrong" sign in the gradient term for $\Phi$ can be thought in analogy to the negative squared mass term of the standard Mexican hat potential. Thus we say that \eqref{lagra} features a ``gradient Mexican hat" for $\Phi$ \cite{Musso:2018wbv,Musso:2019kii}.

The equations of motion are given by
\begin{align}\label{EOM1}
 \partial_t^2 \Phi
 + A\,  \partial_i^2 \Phi
 - 2B\, \partial_i \left(\frac{\partial_i \Phi}{\Xi^6} \partial_j \Phi^* \partial_j \Phi \right)
 &= 0 \ , \\ \label{EOM2}
 \partial_t^2 \Xi
 - \partial_i^2 \Xi
 - \frac{6}{\Xi} \left[B \frac{\left(\partial_i \Phi^* \partial_i \Phi\right)^2}{\Xi^6}
 - H \Xi^6\right]
 &= 0 \ .
\end{align}

\subsection{Ground states}
\label{GroundStatesSection}

There is a large class of possible ground states that break spontaneously translation invariance with different patterns, but it is strongly restricted if we demand that the effective action for perturbations around the ground state is homogeneous, leaving just two possible types (see Appendix  \ref{HomogeneousVacua}). Following the symmetry breaking pattern they exhibit we dub the first ``helical superfluid'' and the second ``meta-fluid''. We will discuss both, pointing out the similarities and differences between the two types of ground states.

\begin{itemize}
    \item {\bf Helical superfluid:}\\
We consider the following static ansatz for the solutions
\begin{align}\label{ansa1}
 \Phi(t,x,y) &= \rho\, e^{i k x}\ ,\\ \label{ansa2}
 \Xi(t,x,y) &= v\ ,
\end{align}
where the compensator field is spatially constant, while the complex field configuration corresponds to a plane-wave of amplitude $\rho$ and wave-vector $k$. All the parameters in the ansatz, $\rho$, $k$ and $v$, are assumed to be non zero, and without loss of generality also real and positive.

The equations of motion descending from \eqref{lagra}, when considered upon the ansatz \eqref{ansa1} and \eqref{ansa2} reduce to 
\begin{align}\label{equazio1}
 \rho^2 k^2\left({2 B k^2 \rho^2}-A{v^6}\right) &= 0\ ,\\ \label{equazio2}
 B k^4 \rho^4 - H v^{12} &= 0\ .
\end{align} 

We can rewrite \eqref{equazio1} and \eqref{equazio2} as follows:
\begin{align}\label{eom1a}
 A &= 2 B \xi \ , \\ \label{eom2a}
 H &=  B \xi^2\ ,
\end{align}
where we have introduced the dimensionless combination
\begin{equation}\label{para}
 \xi = \frac{k^2 \rho ^2}{v^6}=\frac{A}{2B}=\sqrt{\frac{H}{B}}\ ,
\end{equation}
which parameterizes the space of non-trivial static solutions.
Positivity (and reality) of $\xi$ implies $AB > 0$ and $HB > 0$.
This is indeed satisfied by our choice of taking $A$, $B$ and $H$ all positive. Consistency of all the relations in \eqref{para} requires the following relation on the Lagrangian coefficients
\begin{equation}\label{eom12}
 H=\frac{A^2}{4B}\ ,
\end{equation}
necessary to have non-trivial solutions, \emph{i.e.}  $v\neq0$, $k\neq0$ and $\rho\neq 0$; notice that this amounts to a \emph{fine-tuning}. The significance of the fine-tuning becomes apparent when looking at the energy density for a static configuration. For \eqref{eom12} it takes the form of a complete square
\begin{equation}
    \varepsilon=B\Xi^{-6}\left( \partial_i \Phi^* \partial_i \Phi-\frac{A}{2B}\Xi^6 \right)^2=Bv^{-6}\left(k^2\rho^2-\xi v^6\right)^2.
\end{equation}
When evaluated on \eqref{para}, the energy density is zero, so these are minimal energy solutions.

It is easy to see that there are two directions of marginal stability;
in fact we are fixing only the combination $\xi$ given in \eqref{para},
but the ansatz \eqref{ansa1} and \eqref{ansa2} has three independent parameters. In other words, we have a two-dimensional space of ground states for this particular ansatz.

We will expand for small fluctuations around this ground state using the parameterization
\begin{align}\label{flu1in}
 \Phi(t,x,y) &= \rho\, e^{i k x} \left[1 + \phi(t,x,y)\right]
 = \rho\, e^{i k x} \left[1 + \sigma(t,x,y) + i \chi(t,x,y)\right]\ ,\\ \label{flu2in}
 \Xi(t,x,y) &= v\, \left[1 + \tau(t,x,y)\right]\ .
\end{align}

\item {\bf Meta-fluid:}\\
We still consider model \eqref{lagra}, but with a different background ansatz, namely
\begin{align}\label{plameta}
 \Phi &= b\, (x + i y)\ , \\
 \Xi &= v\ , \label{plameta2}
\end{align}
where $b$ and $v$ are respectively a complex and a real constant. In principle there can be more complicated solutions of this type where one introduces two complex constants $b_x$ and $b_y$ such that $\Phi=b_xx+b_y y$. The main difference with the case we study is that \eqref{plameta} keeps a combination of spatial and phase rotations of the complex field unbroken, while the more general solution does not. Since we are mainly interested in the breaking of translation symmetry we keep to the isotropic case in order to avoid further complications.

The equation of motion \eqref{EOM1} for $\Phi$ is automatically solved by the ansatz, while that for $\Xi$, \eqref{EOM2}, eventually leads to
\begin{equation}\label{v}
 v^6 = \frac{4B}{A} |b|^2 \ ,
\end{equation}
where we have used the condition on the coefficients \eqref{eom12}. This guarantees that the energy density of the configuration vanishes, so these are also minimal energy solutions of the same model.
We will perform an expansion of small fluctuations around the background 
\begin{align}\label{eq:metaansatz}
    \Phi &=b (x+iy)+b\left[u_x(t,x,y)+i u_y(t,x,y)\right],\\ 
    \Xi &=v+\tau(x,y,z)\ .
\end{align}
The fluctuations $u_i$ can be interpreted as displacement fields in a solid, in the spirit of the effective actions proposed in \cite{Leutwyler:1996er,Son:2005ak, Nicolis:2013lma}.

\end{itemize}

Finally, it is worth mentioning that there is not really an unbroken phase, even for $A<0$. Indeed the compensator field $\Xi$ appears in the denominator in the interaction term with coefficient $B$ in \eqref{lagra}, and hence the limit $v\to 0$ is not well-behaved. We henceforth always keep $v>0$.

\subsection{Symmetries and Ward-Takahashi identities}
\label{WISection}
 
The action defined by the Lagrangian \eqref{lagra} presents the following symmetries
\begin{itemize}
\item $U(1)$ symmetry: 
\begin{equation}\label{eq:u1symmetry}
    \Phi \rightarrow e^{i \alpha} \Phi \ , \qquad \qquad \Xi \rightarrow \Xi \ ,
\end{equation}
\item Complex shift symmetry:
\begin{equation}\label{eq:complexshift}
    \Phi \rightarrow \Phi + a_R + i\, a_I \ , \qquad \qquad \Xi \rightarrow \Xi \ ,
\end{equation}
\item Dilatation symmetry:
\begin{equation}\label{eq:scaling}
x^\mu\rightarrow e^{-\eta} x^\mu, \qquad
    \Phi \rightarrow  e^{\eta/2}\Phi \ , \qquad \qquad \Xi \rightarrow e^{\eta/2}\Xi \ ,
\end{equation}
\end{itemize}
Note that the $U(1)$ and complex shift symmetries are not independent, we can always use a $U(1)$ transformation to rotate a complex shift into a real one. The set of independent symmetries we discuss will then be dilatations and either the $U(1)$ and real shift or the complex shifts. 

In the helical state the $U(1)$ symmetry is broken together with translations along the $x$ direction to a diagonal combination. Real shifts and dilatations are also broken. The symmetry breaking pattern is quite different in the meta-fluid. In this case, it is the complex shift symmetry the one broken with translations, in both $x$ and $y$ directions, to a diagonal combination. A $U(1)$ symmetry that combines the phase change of the complex field and spatial rotations survives, so this phase is rotationally invariant. As in the previous case, dilatation symmetry is also broken.

The na\"{\i}ve counting of Nambu-Goldstone bosons would give us three gapless modes in each case: the Nambu-Goldstone modes associated to $U(1)$, real shift and dilatations in the helical state and the Nambu-Goldstone modes associated to real and imaginary shifts and dilatations in the meta-fluid state. As we will see the na\"{\i}ve counting fails and a mode becomes gapped. We will return to the issue of this counting in Section \ref{sec:count}.

In the meta-fluid state the identification of the fluctuations is more or less evident, $u^i$ should be associated to spatial translations/complex shifts while $\tau$ should correspond to scale transformations. In the helical state $\chi$ is clearly related to $U(1)$ rotations/translations in the $x$ direction, but the role of $\sigma$ and $\tau$ is not so obvious. In order to help with the identification of the modes in the following we will consider the Ward-Takahashi identities associated to symmetries. A more detailed derivation of the identities can be found in Appendix~\ref{sec:WT}.

The Ward-Takahashi identities at linear order in the fluctuations return different combinations of the linear equations of motion that we will obtain from the Lagrangian in \eqref{prim}, \eqref{segu} and \eqref{terc}.
The extra information we get from the Ward-Takahashi identities is that, when considering the decoupling or high momentum limit  (which we will implement by formally taking $k\rightarrow 0$, though of course we keep the premise that $k\neq 0$ for symmetry breaking to happen), one can establish a connection between the fluctuation fields $\chi$, $\sigma$ and $\tau$ and the $U(1)$, the real shift and the dilatation symmetries. Similarly, for the meta fluid one can identify the dispersion relations that correspond to each mode at high momentum. 
Note that in the perspective where \eqref{lagra} is already an effective theory, the dispersion relations at high momentum would in principle be modified by putative higher derivative terms not included in the Mexican hat model we are studying (scale invariance would be explicitly broken by such corrections). However those would come suppressed by a mass scale that we assume to be much larger than any of the scales in the model so it is still sensible to discuss a high momentum limit. 
\\

\paragraph{ \bf $U(1)$ symmetry} ~\\

\noindent
The $U(1)$ current corresponding to the Lagrangian $\partial_\mu \Phi^* \partial^\mu \Phi$ has the form
\begin{equation}
 j_\mu = \frac{i}{2} \left(
 \Phi \partial_\mu \Phi^*
 -\Phi^* \partial_\mu \Phi
 \right)\ .
\end{equation}
Thus, for the model \eqref {lagra} we have
\begin{align}
 J_0 &= j_0\ ,\\
 J_i &= -\left(A-2B\frac{\partial_j \Phi^* \partial_j \Phi}{\Xi^6}\right)j_i\ ,
\end{align}
whose conservation is encoded in the continuity equation\footnote{In our conventions $\partial^\mu=(\partial_t,-\partial_i)$.}
\begin{equation}
 \partial^\mu J_\mu = 0\ .
\end{equation}
Expanding to linear order in the fluctuations of the helical superfluid we have
\begin{equation}\label{con_cor}
\partial_t^2 \chi-2A \partial_x\left[k(\sigma-3\tau)+\partial_x\chi\right]=0\ ,
\end{equation}
In the $k\rightarrow 0$ limit one finds
\begin{equation} \label{U1kLimit}
 \partial_t^2 \chi-2A \partial_x^2 \chi\simeq  0\ ,
\end{equation}
indicating that at large frequency and momentum compared to $k$, the  perturbation $\chi$ maps to the Nambu-Goldstone boson of the $U(1)$ symmetry, with a dispersion relation
\begin{equation}
\omega^2\simeq 2A q_x^2, \qquad q_x \gg k. \label{eq:displineon}
\end{equation}
This mode has an unusual dispersion relation, and we will refer to it as a `lineon' since it moves on a line. We will discuss this in more detail when we introduce the connection to fractons.

\paragraph{Shift symmetry} ~\\

\noindent
The (complex) shift current corresponding to the Lagrangian $\partial_\mu \Phi^* \partial^\mu \Phi$ is given by
\begin{equation}\label{shi_sta}
 j_\mu^{(s)} = \partial_\mu \Phi
 \ ,
\end{equation}
where the $s$ label stands for ``shift''. The current is linear in the field because the field variation is a constant. 
Using \eqref{shi_sta}, the current for the model \eqref{lagra} can be expressed as follows:
\begin{align}\label{eq:shiftcurrent}
 J_0^{(s)} &= j_0^{(s)}\ ,\\
 J_i^{(s)} &= -\left(A-2B\frac{\partial_j \Phi^* \partial_j \Phi}{\Xi^6}\right)j_i^{(s)}\ .
\end{align}
The associated continuity equation is
\begin{equation}
 \partial^\mu J_\mu^{(s)} = 0\ ,
\end{equation}
which, at linear level in the fluctuations of the helical superfluid, gives two linearly independent equations, \eqref{con_cor} and
\begin{align}\label{con_shi_1}
\partial_t^2 \sigma+2Ak \left[k(\sigma-3\tau)+\partial_x \chi\right] &=0 \ .
\end{align}
In the $k\rightarrow 0$ limit, we get
\begin{align}
 \partial_t^2 \sigma &\simeq 0\ . \label{U1kLimitAgain}
\end{align}
Therefore, at large frequencies and momenta compared to $k$, the perturbation $\sigma$ can be identified with the Nambu-Goldstone mode of (real) shifts. Again, the unusual dispersion relation $\omega^2\simeq 0$ will be discussed later on.

For the meta-fluid it is convenient to study only the Ward-Takahashi identity of complex shifts. 
To linear order in the fluctuations the conservation of the complex shift current produces the equations
\begin{equation}
    v\left(\partial_t^2 u_i-A\partial_i \partial_k u_k\right)+6A\partial_i\tau=0\ .
\end{equation}
At high momentum, $\tau$ is decoupled and the displacements $u_i$ combine in two modes with dispersion relations
\begin{equation}
    \omega^2\simeq 0\ ,\qquad \ \omega^2\simeq  A (q_x^2+q_y^2)\ ,
\end{equation}
where the trivial mode corresponds to the transverse component $\partial_k u_k=0$ and the propagating mode to the longitudinal component.

\paragraph{Dilatation symmetry} ~\\

The Lagrangian \eqref{lagra} being scale invariant ensures us that we can improve the energy-momentum tensor such that the dilatation conserved current takes the form 
\begin{equation}
 D^\mu = {\cal T}^\mu_{\ \ \nu} x^\nu - V^\mu \ ,
\end{equation}
where $ V^\mu$ is called the virial current. Therefore, the conservation equation
\begin{equation}
 \partial_\mu D^\mu = 0
\end{equation}
is equivalent to say that, on-shell, the trace of the improved energy-momentum tensor is zero up to a total divergence of the virial current
\begin{equation}
 {\cal T}^\mu_{\ \ \mu} = \partial^\mu V_\mu \ .
 \label{dilatationconservation}
\end{equation}
The improved energy-momentum tensor contains the following terms
\begin{equation}
 {\cal T}^\mu_{\ \ \nu} \equiv T^\mu_{\ \nu} + \left(\Box \delta^\mu_\nu - \partial^\mu \partial_\nu\right)\left(\frac{1}{4}\, |\Phi|^2 + \frac{1}{8}\, \Xi^2\right)
 + \frac{A+1}{2}\, \theta^\mu_{\ \nu} \label{improvedTmunu}
\end{equation}
where
\begin{align}
 T^\mu_{\ \nu} & = \frac{\delta {\cal L}}{\delta \partial_\mu X^I} \partial_\nu X^I 
 -\delta^\mu_\nu {\cal L}\ , \\
 \theta^i_{\ j} & \equiv \left(\partial_k^2 \delta_{ij} - \partial_i\partial_j\right) |\Phi|^2  \ .
\end{align} 
By injecting the equations of motion in the trace of \eqref{improvedTmunu}, we have that the virial current is given by
\begin{align}
 V_0 &= 0\ , \\
 V_i &= \frac{B}{\Xi^6} \left(\partial_k\Phi^* \partial_k\Phi\right) \partial_i |\Phi|^2 \ .
\end{align}
We now have an explicit expression for \eqref{dilatationconservation}, which at linear order in the fluctuations of the helical superfluid gives
\begin{equation}
    v^2(\partial_i^2\tau-\partial_t^2\tau)=2\rho^2 \partial_t^2\sigma+8 k \rho^2 A \left(k(3\tau-\sigma)-\partial_x\chi \right).
\end{equation}
In the $k\rightarrow 0$ limit (assuming $v$, $\rho$ can be kept fixed), one obtains
\begin{equation}\label{eq:dispdilaton}
 \partial_i\partial^i\tau -\partial_t^2\tau=0\ ,
\end{equation}
where we have used \eqref{U1kLimitAgain}. Then, for large values of frequency and momenta, $\tau$ can be identified with the Nambu-Goldstone boson for dilatations. In this case the dispersion relation is the usual one for a relativistic massless mode
\begin{equation}
    \omega^2 \simeq q_x^2+q_y^2\ . \label{eq:largeMomentRel}
\end{equation}

For the meta-fluid, the dilatation Ward-Takahashi identity produces the following equation
\begin{equation}
    \left(\partial_i^2-\partial_t^2\right)\tau=\frac{12 A |b|^2}{v^2}\left(6\tau-v\partial_k u^k \right).
\end{equation}
At high momentum the displacement fields decouple and the dilaton has a relativistic dispersion relation \eqref{eq:largeMomentRel} as in the helical superfluid.

\subsection{Connection to fractons}
\label{fracchia}

The unusual dispersion relations we have found in \eqref{eq:displineon} and \eqref{U1kLimitAgain} are not just a peculiarity of the decoupling limit but they are also observed at small frequency and momentum, as we will show in the next sections. A possible way to understand their origin is through emergent symmetries of linearized perturbations around the translation-breaking ground states. These symmetries involve coordinate-dependent shifts of the fields similar to those introduced in fracton models \cite{Pretko:2020cko,nandkishore_fractons_2019} and are linked to excitations that are immobile or restricted to move in a subdimensional space.

In order to identify the emergent symmetry more easily, we will proceed by studying the quadratic Lagrangian of the perturbations and integrating out the gapped mode. The resulting effective Lagrangian admits a derivative expansion where the symmetry becomes manifest.

\paragraph{Helical superfluid}~\\

The action to quadratic order in the fluctuations is 
\begin{equation}\label{lquadheli}
    {\cal L}=\frac{v^2}{2}\partial_\mu\tau\partial^\mu \tau+\rho^2(\partial_t \chi)^2+\rho^2(\partial_t\sigma)^2-2A\rho^2\left[\partial_x \chi +k(\sigma-3  \tau) \right]^2.
\end{equation}
In this form, we already observe emergent coordinate-dependent shift symmetries, namely
\begin{equation}\label{eq:fractonsym}
    \delta\chi=\alpha(y)+\beta(x,y), \ \ \delta\sigma=-\frac{1}{k}\partial_x\beta(x,y)+3\delta+3\gamma_i x^i, \ \ \delta\tau(x,y)=\delta+\gamma_i x^i.
\end{equation}
The emergence of these symmetries may explain in part the fractonic behavior observed from the analysis of the Ward-Takahashi identities. The dilaton $\tau$ has the symmetry of a massless field, the symmetry under the transformation $\beta$ can be used to introduce an arbitrary dependence on both $x$ and $y$ in $\sigma$ while the remaining transformation $\alpha$ allows an arbitrary dependence on $y$ in $\chi$. In this way, the identification of $\sigma$ as a fracton and $\chi$ as a lineon appears naturally. Note that these are not symmetries of the full action, so it is expected that terms of higher order in the fluctuations will not be invariant under them, however this only affects indirectly the dispersion relations by radiative corrections.

We can diagonalize the mass terms by performing a rotation of the fields
\begin{equation}\label{eq:massrot}
\left(\begin{array}{c} 
v\tau\\  \sqrt{2}\, \rho \sigma 
 \end{array}\right)
=\left(\begin{array}{cc} 
\cos\theta &  \sin\theta \\
-\sin\theta & \cos\theta
 \end{array}\right)
\left(\begin{array}{c} 
\eta \\  \varphi 
 \end{array}\right)\ ,
\end{equation}
by an angle
\begin{equation}\label{eq:tanth}
    \tan\theta=\frac{v}{3\sqrt{2}\, \rho}\ .
\end{equation}
The action becomes
\begin{equation}
    {\cal L}=\rho^2(\partial_t \chi)^2+\frac{1}{2}(\partial_t\varphi)^2+\frac{1}{2}(\partial_t\eta)^2-\frac{1}{2}[\partial_i(\cos\theta \eta+\sin\theta \varphi)]^2-2A\rho^2\left[\partial_x \chi -\frac{m_\eta}{2\sqrt{A}\rho} \eta\right]^2.
\end{equation}
The mass of $\eta$ equals to
\begin{equation}
    m_\eta^2=\frac{36A k^2\rho^2}{v^2}\left(1+\frac{v^2}{18\rho^2}\right)=2 Ak^2\left(1+18\frac{\rho^2}{v^2}\right).
    \label{massEFTmeta}
\end{equation}
We can now group terms linear in $\eta$ inside a squared term (after integrating by parts) and subtract the appropriate $\eta$-independent terms
\begin{equation}
\begin{split}
    {\cal L}=&\frac{1}{2}(\partial_t\eta)^2-\frac{1}{2}\cos\theta^2(\partial_i\eta)^2-2A\rho^2\left[\partial_x \chi+\frac{\sin\theta\cos\theta}{2m_\eta\sqrt{A}\rho}\partial_i^2\varphi -\frac{m_\eta}{2\sqrt{A}\rho} \eta\right]^2\\
    &+\rho^2(\partial_t\chi)^2+\frac{1}{2}(\partial_t\varphi)^2-\frac{1}{2}\sin^2\theta(\partial_i\varphi)^2\\
    &+\frac{2\sqrt{A}\rho}{m_\eta}\sin\theta\cos\theta\partial_x \chi\partial_i^2\varphi+\frac{\sin^2\theta\cos^2\theta}{2m_\eta^2}(\partial_i^2\varphi)^2 \ .
\end{split}
\end{equation}
Next, we integrate out $\eta$ expanding its solution in derivatives, starting at lowest order with
\begin{equation}
    \eta\simeq \frac{2\sqrt{A}\rho}{m_\eta}\left(\partial_x \chi +\frac{\sin\theta\cos\theta}{2m_\eta\sqrt{A}\rho}\partial_i^2\varphi\right).
\end{equation}
Then, up to the fourth order in derivatives, we get
\begin{equation}\label{lag_chi}
\begin{split}
    {\cal L}=&\rho^2(\partial_t\chi)^2+\frac{1}{2}(\partial_t\varphi)^2-\frac{1}{2}\sin^2\theta(\partial_i\varphi)^2+\frac{2\sqrt{A}\rho}{m_\eta}\sin\theta\cos\theta\partial_x \chi\partial_i^2\varphi\\ 
    &+\frac{2A\rho^2}{m_\eta^2}\left[(\partial_t\partial_x\chi)^2-\cos^2\theta (\partial_i\partial_x\chi)^2 \right]+\frac{\sin^2\theta\cos^2\theta}{2m_\eta^2}(\partial_i^2\varphi)^2\ .
\end{split}
\end{equation}
Both $\chi$ and $\varphi$ are gapless and have constant shift symmetries so there are corresponding conserved charges. Furthermore, up to total derivatives in the Lagrangian \eqref{lag_chi}, $\chi$ can be shifted by a term depending on the coordinates
\begin{equation}\label{chi_shi}
\chi\to \chi+a_i x^i+c_{ij}x^i x^j+ f(y)\ .
\end{equation}
Symmetry under shifts by linear terms imply that the dipole moment of the charge is conserved, while shifts under quadratic terms imply the conservation of quadrupole and second radial moment. This is characteristic of models of fractons that are immobile. Although higher derivative terms might spoil the shift symmetries, this would only affect the dispersion relations at higher order.

To quadratic order in momentum, the dispersion relations of the gapless fluctuations are
\begin{equation}\label{eq:helicalgapless}
    \begin{split}
        \omega_\chi^2\simeq &\, 0\ ,\\
        \omega_\varphi^2\simeq &\, \sin^2\theta\, q_i^2=\frac{v^2}{18\rho^2+v^2}q_i^2\ .
    \end{split}
\end{equation}

\paragraph{Meta-fluid}~\\
\label{MetafluidFracton}

To linear order, the spatial derivatives of $\Phi$  are
\begin{equation}
    \partial_i \Phi=b(\delta_i^x+i\delta_i^y)+b(\partial_i u_x+i\partial_i u_y)\ \quad \Rightarrow \quad \partial_i \Phi^*\partial_i \Phi =|b|^2\left(2+2\partial_i u_i+(\partial_i u_j)^2 \right)\ .
\end{equation}
Then, expanding the action to quadratic order in the fields, we find
\begin{equation}\label{eq:elastic1}
    {\cal L}=\frac{1}{2}\partial_\mu\tau \partial^\mu\tau +|b|^2\partial_t u_i \partial_t u_i-A|b|^2\left(\partial_i u_i-\frac6v \tau \right)^2\ .
\end{equation}
We can also write this action in the following way
\begin{equation}
    {\cal L}=\frac{1}{2}\partial_\mu\tau \partial^\mu\tau-\frac{1}{2}m_\tau^2 \tau^2 +|b|^2\left(\partial_t u_i \partial_t u_i-C^{ijkl}\partial_i u_j \partial_k u_l+\frac{12 K}{v} \tau \partial_i u_i\right)\ .
\end{equation}
The coefficients $C^{ijkl}$ are the components of the elasticity tensor, that in this case only has a bulk component
\begin{equation}
    C^{ijkl}=K\delta^{ij}\delta^{kl},\ \qquad K=A\ ,
\end{equation}
with $K$ the bulk modulus, which also enters in the coupling between the dilaton and the bulk strain. The mass of the dilaton is
\begin{equation}
    m_\tau^2=72 K\frac{|b|^2}{v^2}\ .
    \label{metaFluidMass}
\end{equation}
A large $v$ limit would make the dilaton massless and decoupled from the elastic theory at low energies, this latter remaining otherwise unaffected. Roughly, if there is a big hierarchy between the spontaneous breaking of dilatations and that of translations, one does not expect the low-energy elastic theory to be sensitive to the dilaton physics.

Since the shear modulus vanishes, any deformation with $\tau=0$, $\partial_i u_i=0$ has zero energy. Then, the elastic part describes a fluid or a meta-fluid. Note that constant changes in volume can be compensated with a shift of the dilaton, so scale invariance is preserved in this sense. This implies that there is a zero mode associated to the dilatation symmetry and a massive mode which corresponds to the combination squared in \eqref{eq:elastic1}.

We can separate the gapped and gapless modes by doing the shift
\begin{equation}
    \tau\to \sigma+\frac{v}{6}\, \partial_i u_i\ .
\end{equation}
Then
\begin{equation}
    {\cal L}=\frac{1}{2}\partial_\mu\sigma \partial^\mu\sigma -\frac{1}{2}m_\tau^2 \sigma^2+\frac{v}{6}\partial_\mu \sigma \partial^\mu \partial_i u_i +|b|^2\partial_t u_i \partial_t u_i+\frac{v^2}{72}\partial_\mu \partial_i u_i \partial^\mu\partial_j u_j\ .
\end{equation}
We can further complete the square
\begin{equation}
 \begin{split}
    {\cal L}&=\frac{1}{2}\partial_\mu\sigma \partial^\mu\sigma -\frac{1}{2}m_\tau^2 \left(\sigma+\frac{v}{6 m_\tau^2} \partial_\mu\partial^\mu \partial_i u_i\right)^2 +|b|^2\partial_t u_i \partial_t u_i\\
    & \qquad +\frac{v^2}{72}\partial_\mu \partial_i u_i \partial^\mu\partial_j u_j+\frac{v^2}{72 m_\tau^2}\left(\partial_\mu\partial^\mu \partial_i u_i \right)^2\ .
 \end{split}
\end{equation}
Integrating out $\sigma$ implies solving order by order in derivatives with the leading term
\begin{equation}
    \sigma \simeq -\frac{v}{6m_\tau^2}\partial_\mu \partial^\mu \partial_i u_i\ .
\end{equation}
To sixth order in derivatives in the action, we are left with
\begin{equation}
    {\cal L}\simeq |b|^2\partial_t u_i \partial_t u_i+\frac{v^2}{72}\partial_\mu \partial_i u_i \partial^\mu\partial_j u_j+\frac{v^2}{72 m_\tau^2}\left(\partial_\mu\partial^\mu \partial_i u_i \right)^2\ .
    \label{EFTmetaGapless}
\end{equation}
In this form, we also observe that the shear strain has zero energy and that the action is symmetric under constant changes of the bulk strain. This implies that there are linear and quadratic shift symmetries
\begin{equation}\label{eq:fractusym}
    \delta u_i=a_i+b_{ij} x^j+c_{ijk}x^j x^k\ .
\end{equation}
Then, we have that, not only the charges associated to the constant shifts, but also their dipole and second moments are conserved, this is characteristic of fractonic models. The larger symmetry associated to arbitrary shear and rotational strains corresponds to transverse transformations
\begin{equation}\label{eq:rotshear}
    \delta u_i= \epsilon_{ik}\partial_k\omega({\bm x})+\left(\partial_i\partial_j-\delta_{ij}\partial_k^2\right) V^j({\bm x})\ .
\end{equation}

\section{Dispersion relations}\label{sec:dispersion}

By a standard pertubation analysis of the model \eqref{lagra} around the respective backgrounds of the helical superfluid and the meta-fluid, we will compute the dispersion relations of the fluctuations. This will support and refine some of the results and interpretations which we already derived in the preceding sections.

\subsection{Helical superfluid}
\label{DispRelHelicalSection}

As stated in Subsection \ref{GroundStatesSection}, we perform a fluctuation of the model \eqref{lagra} around a plane-wave background where we consider the parameterization
\begin{align}\label{flu1}
 \Phi(t,x,y) &= \rho\, e^{i k x} \left[1 + \phi(t,x,y)\right]
 = \rho\, e^{i k x} \left[1 + \sigma(t,x,y) + i \chi(t,x,y)\right]\ ,\\ \label{flu2}
 \Xi(t,x,y) &= v\, \left[1 + \tau(t,x,y)\right]\ .
\end{align}
The equations of motion at linear order for the fluctuations are%
\footnote{We remind the reader that the parameters $k$, $\rho$ and $v$ are not independent, but related by \eqref{para}. We will refrain from expressing one of the parameters in terms of the others, but instead aim at writing the various expressions in their simplest form, here and in the rest of the paper.} 
\begin{align}\label{prim}
 2 A (k + i \partial_x) \left[k (\sigma -3 \tau )+ \partial_x \chi\right] + \partial_t^2 (\sigma -i \chi ) &=0\ ,\\ \label{segu}
 2 A (k - i \partial_x) \left[k (\sigma -3 \tau )+ \partial_x \chi\right] + \partial_t^2 (\sigma +i \chi ) &=0\ ,\\ \label{terc}
 12 A k \rho^2 \left[k (\sigma -3 \tau )+ \partial_x\chi \right]-v^2 \left(-\partial_x^2 - \partial_y^2 + \partial_t^2\right) \tau &= 0\ .
\end{align}
Going to Fourier space, we obtain a homogeneous algebraic system determined by the kinetic matrix:

\begin{equation}\label{mat}
 M = \rho^2 \left(
\begin{array}{ccc}
 \omega^2-2 A k^2 & -2 i A k q_x  & 6 A k^2 \\
 2 i A k q_x & \omega^2-2 A q_x^2 & -6 i A k q_x \\
 6 A k^2 & 6 i A k q_x & \frac{1}{2}\left(\omega^2-q_x^2-q_y^2\right) \frac{v^2}{\rho^2} -18 A k^2 
\end{array}
\right)\ ,
\end{equation}
where the first row corresponds to $\sigma$, the second one to $\chi$ and the third one to $\tau$. 

In order to have non-trivial solutions for the fluctuations, the determinant for the kinetic matrix should vanish, 
\begin{equation} 
 \text{det}(M) = \frac{\omega^2 \rho^4}{2} \left\{ v^2 \left(\omega^2-q_x^2-q_y^2\right)
   \left[\omega^2-2 A \left(k^2+q_x^2\right)\right]-36 A k^2 \omega^2 \rho^2\right\}=0\,.
   \label{detMat}
\end{equation}
This leads to a set of conditions for the frequency and momenta that determine the dispersion relations.
The fluctuation determinant \eqref{detMat} has a $\omega^2$ factor, producing a gapless mode whose dispersion 
relation is trivial, \emph{i.e.} identically zero, $\omega=0$. 
Apart from such a trivial mode, the spectrum features a gapless and a gapped mode
\begin{align}
 m_2^2 &= 0\ ,\\ \label{gap3}
 m_3^2 &= 2 A k^2 \left(1+18 \frac{\rho^2}{v^2}\right)\ . 
\end{align}
Let us remark that the mass \eqref{gap3} agrees with the mass coming from the effective field theory analysis \eqref{massEFTmeta}, $m_3\equiv m_\eta$.

Proceeding to compute the dispersion relations, we obtain:
\begin{align}
 \omega_1^2 &= 0\ , \label{zerozero} \\
  \omega^2_{2,3} &=\frac12\left[ 2 A  q_x^2+ q_x^2+q_y^2+m_\eta^2\mp\sqrt{\Delta}\right] \ , \label{mode23}
\end{align}
with
\begin{equation}
    \Delta = \left\{2 A q_x^2+q_x^2+q_y^2+m_\eta^2\right\}^2
   -8 A  \left(k^2+q_x^2\right) \left(q_x^2+q_y^2\right)\ .
\end{equation}
The expansion at low momenta provides
\begin{align}
 \omega_1^2 &= 0\ , \label{fracton} \\
 \omega_2^2 &= \frac{v^2}{18\rho^2+v^2} \left(q_x^2 + q_y^2\right) + {\cal O}(q^4)\ ,  \label{gapless}\\
 \omega_3^2 &= m_\eta^2
 +2Aq_x^2+\frac{18\rho^2}{18\rho^2+v^2}\left(q_x^2 + q_y^2\right)+{\cal O}\left(q^4\right) \ .\label{gapped}
\end{align}
To recover the results obtained from the Ward-Takahashi identities of Section \ref{WISection}, and in particular the fractonic behavior, we consider the large-momentum behavior of \eqref{mode23}. Of course, the exact trivial mode will remain trivial in any $q$ limit. In order to take the large-momentum limit, we simply take $q_x,q_y\gg m_\eta$. Then, we find
\begin{equation}\label{eq:om2}
    \omega_2^2 
    \simeq \left\{\begin{array}{ccc}
         q_x^2+q_y^2   & \text{if} & (2A-1)q_x^2-q_y^2>0\\
         2A q_x^2  & \text{if} & (2A-1)q_x^2-q_y^2<0
    \end{array}\right. \ ,
\end{equation}
and
\begin{equation}\label{eq:om3}
     \omega_3^2 
    \simeq \left\{\begin{array}{ccc}
         2A q_x^2   & \text{if} & (2A-1)q_x^2-q_y^2>0\\
          q_x^2+q_y^2  & \text{if} & (2A-1)q_x^2-q_y^2<0
    \end{array}\right. \ .
\end{equation}
Thus, $\omega_2$ and $\omega_3$ swap their roles depending on the sign of $(2A-1)q_x^2-q_y^2$, which in general depends on the direction in the momentum plane. Note that if $A\leq 1/2$ this quantity is always negative, so in that case $\omega_2$ and $\omega_3$ do not change with direction.

\subsection{Identification of the modes}\label{sec:idenmod}

In order to study the Nambu-Goldstone nature of \eqref{fracton}, \eqref{gapless} and \eqref{gapped}, we need to determine how they relate to a local spacetime modulation of the various symmetry-originated zero modes. The study of the Ward-Takahashi identities gave us already a glance into such associations at large momentum. At low momentum, instead, one can get useful information from the effective action, which we have already derived to establish the connection to fractons. Accordingly, we will identify $\chi$ with the $U(1)$ Nambu-Goldstone mode and $\tau$ and $\sigma$ with the dilaton and {\em shifton} respectively.

\begin{enumerate}
    \item Low momentum: \\
    Comparing \eqref{fracton}, \eqref{gapless} and \eqref{gapped} with \eqref{eq:helicalgapless} and \eqref{massEFTmeta}, we can make the following identifications
    \begin{enumerate}[i]
        \item Trivial mode: $\omega_1$, mostly $\chi$.
        \item Gapless mode: $\omega_2$, mixture of $\tau$ and $\sigma$. According to \eqref{eq:massrot} and \eqref{eq:tanth} if $v\gg \rho$ ($k/\rho^2\gg 1$) it would be mostly $\tau$ and if $v\ll \rho$ ($k/\rho^2\ll 1$) it would be mostly $\sigma$.
        \item Gapped mode: $\omega_3$, mixture of $\tau$ and $\sigma$ orthogonal to the gapless mode.
    \end{enumerate}
    \item High momentum:\\
    Comparing \eqref{U1kLimitAgain}, \eqref{eq:displineon} and \eqref{eq:dispdilaton} with \eqref{eq:om2} and \eqref{eq:om3}, we can identify
    \begin{enumerate}[i]
        \item Trivial mode: $\omega_1$ mostly $\sigma$.
        \item Lineon: $\omega_3$ (for $(2A-1)q_x^2-q_y^2>0$) or $\omega_2$ (for  $(2A-1)q_x^2-q_y^2<0$), mostly $\chi$.
        \item Relativistic mode:  $\omega_2$ (for $(2A-1)q_x^2-q_y^2>0$) or $\omega_3$ (for  $(2A-1)q_x^2-q_y^2<0$),  mostly $\tau$ .
    \end{enumerate}
\end{enumerate}

These identifications unveil a strong change in the nature of the modes as a function of momentum, this being a reflection of the mixing induced by the breaking of translation symmetry. For $A>1/2$, the transmutation does not only occur in the transition from low to large momentum but also depending on the direction in the momentum plane.

We would like now to pause a moment to comment on the relation with the non-scale invariant model of \cite{Musso:2018wbv}, where precisely the helical ground state was considered. The model is basically the same as the present one, where however the fluctuation $\tau$ is frozen. The spectrum is easily obtained from the determinant of the upper-left 2-by-2 submatrix of \eqref{mat}. It consists of a trivial fractonic mode, and a gapped lineon. Hence we see that the compensator field enforcing scale invariance is a highly non-trivial addition to the model, yielding non-trivial mixing among the modes, and their identification.

\subsection{Meta-fluid}

In order to get the dispersion relations for the meta-fluid, we proceed in a similar fashion as for the helical superfluid. By referring to Section \ref{GroundStatesSection}, we study the fluctuations of the model \eqref{lagra} around the background given in \eqref{plameta} and \eqref{plameta2} with the following parametrization:
\begin{align}
 \Phi(t,x,y) &= b \Big[x+i y+ u_x(t,x,y) + i u_y(t,x,y)\Big]\ ,\\
 \Xi(t,x,y) &= v + \tau(t,x,y)\ .
\end{align}
At first order in the fluctuations, the equations of motion are given by
\begin{align}
 & \partial_t^2 u_x -A \partial_x\left(\partial_xu_x-\frac6v \tau\right)= 0\ ,\\
 &  \partial_t^2 u_y -A \partial_y\left(\partial_yu_y-\frac6v \tau\right) = 0\ , \\
 & \partial_t^2\tau -\partial_i^2\tau +12A\frac{|b|^2}{v}\left(\frac6v \tau-\partial_i u_i\right)= 0\ .
\end{align}
The quadratic fluctuation matrix in Fourier space is
\begin{equation}
M=|b|^2 \left(
\begin{array}{ccc}
 \omega^2-Aq_x^2
 & -Aq_xq_y
 & -6i \frac{A}{v} q_x \\
 -Aq_xq_y & \omega^2-Aq_y^2 
   & -6i \frac{A}{v} q_y \\
 6i \frac{A}{v} q_x
 & 6i \frac{A}{v} q_y & \frac{1}{2|b|^2} \left(\omega^2- q_x^2-q_y^2\right) - 36\frac{ A }{v^2}
\end{array}
\right)\ , \label{KineticMatrixMeta}
\end{equation}
where the two first lines correspond to $u_x$ and $u_y$ while the last line is associated to $\tau$. The determinant of this matrix is given by
\begin{align}\label{detmetafluid}
    \det (M)= \frac{|b|^4 \omega^2}{2}\left[ (\omega^2-q^2)(\omega^2-Aq^2)-72\frac{A|b|^2}{v^2}\omega^2\right]\ ,
\end{align}
where we have used $q^2=q_x^2+q_y^2$. Take notice that the above expression is completely isotropic. 
The dispersion relations, given by the roots of the determinant, are the following
\begin{align}
    \omega_1^2 & = 0\ , \\
    \omega_{2,3}^2 & = \frac12\left\{(1+A)q^2 +m_\tau^2\mp \sqrt{\left[(1+A)q^2 +m_\tau^2\right]^2-4Aq^4}\right\}\ ,
\end{align}
where we have used \eqref{metaFluidMass}.

At low momenta, we find the following dispersion relations:
\begin{align}
 \omega_1^2 & = 0\ ,\label{trivialMeta}\\
 \omega_2^2 &= \frac{A}{m_\tau^2} q^4  + {\cal O}\left(q^6\right)\ , \label{nonTrivialMeta}\\
 \omega_3^2 &= m_\tau^2
 + (1+A)q^2  + {\cal O}\left(q^4\right)\ .  \label{massiveMeta}
\end{align}
We obtain a similar mass spectrum as for the helical superfluid: two massless modes, one being exactly trivial, and a gapped mode. However, we have some qualitative differences in the dispersion relations. Indeed, the non-trivial massless mode has a quadratic dispersion relation while in the helical background it has a linear behavior. This can be traced back to the effective theory for gapless modes \eqref{EFTmetaGapless}, where there are terms with two time derivatives and four spatial derivatives but there are no terms with just two spatial derivatives.

An additional qualitative difference with the helical case is that all the dispersion relations are isotropic in the meta-fluid case. As it was mentioned in Section \ref{WISection}, the meta-fluid background preserves an effective rotation symmetry from the diagonal breaking of $U(1)$ and spatial rotations.  This is actually due to the particular ansatz \eqref{plameta} for the solution. It is possible to choose a more general solution which will lead to anisotropies in the determinant of the kinetic matrix and hence in the spectrum. The other features of the latter would however be unchanged. Hence, we prefer to deal with the isotropic meta-fluid, for a better clarity of the resulting expressions. On the other hand, in the plane wave background spatial rotations are necessarily broken by the choice of a preferred direction in the solution.

At small momentum, the massive mode \eqref{massiveMeta} is associated to the fluctuation $\tau$, as it can be observed in the diagonalization of \eqref{KineticMatrixMeta} in the $q_i\rightarrow 0$ limit. We notice that the association to $\tau$ matches the effective study of Section \ref{MetafluidFracton} and that we recover the mass \eqref{metaFluidMass}.  

For large momenta, the non-trivial modes have the following dispersion relations
\begin{align}
    \omega_{2,3}^2  \simeq \frac{q^2}2\Big[1+A \mp |1-A|\Big]\ ,
\end{align}
so that when $A\leq1$ we have that $\omega_2 \simeq \sqrt{A}|q|$ and $\omega_3\simeq |q|$ (with $|q|=\sqrt{q^2}$), while when $A>1$ we have the opposite, $\omega_2\simeq |q|$ and $\omega_3 \simeq \sqrt{A}|q|$. 
By looking at the kinetic matrix in the large $q$ limit, \emph{i.e.} neglecting all non-leading terms in $\omega$ or $q$, we find that the mode with $\omega\simeq |q|$ is always aligned with $\tau$, while the mode with $\omega\simeq \sqrt{A}|q|$ is aligned with the longitudinal combination of the $u_i$ (the other being always the trivial immobile mode). Hence, when $A>1$, the modes $\omega_2$ and $\omega_3$ switch nature when going from low to high momenta.
On the other hand, note that for the meta-fluid, the trivial mode is always the transverse part of $u_i$, for all momenta.

Again, let us comment briefly on the possibility to have a meta-fluid ground state in the model of \cite{Musso:2018wbv}, where $\tau$ is frozen. In this case, we can easily see that there is no scale in the spectrum. Eventually, the spectrum consists of a trivial fractonic mode, and a gapless isotropic mode with linear dispersion relations. Hence we notice that also in this case, the addition of the compensator field changes quite radically the spectrum, due to non-trivial mixing.

\section{Finite density}\label{OrignModelFiniteDensity}

As it has already been established for internal compact symmetries, working at finite density modifies the spectrum of Nambu-Goldstone modes associated to spontaneous symmetry breaking \cite{Nicolis_2013,Nicolis:2013sga,Watanabe:2013uya,Argurio:2020jcq}. We would like to probe how these results extend in our specific spacetime symmetry breaking pattern. To do so, we switch on a chemical potential $\mu$ for the $U(1)$ symmetry of the theory \eqref{lagra} in the framework of the helical superfluid background. We do not extend the analysis to the meta-fluid because in the presence of a chemical potential the effective action is no longer homogeneous.

The chemical potential introduces a new term in the effective potential $\sim -\mu^2\rho^2$, that makes it unbounded from below and would produce a runaway behavior. Something similar occurs in the model of \cite{Argurio:2020jcq} for the simultaneous breaking of scale invariance and an internal symmetry. In that simpler case the issue was solved by introducing a small deformation of the model that lifts the space of minimal energy states at zero chemical potential and stabilizes it at finite chemical potential. The results at zero density can be recovered by simultaneously sending the chemical potential and the deformation to zero. Following the same reasoning we introduce a new term with coupling $\lambda^2$ that preserves the $U(1)$ and dilatation symmetries
\begin{equation}\label{lagraMod}
 {\cal L} =
 \partial_t \Phi^* \partial_t \Phi
 + A \partial_i \Phi^* \partial_i \Phi
 +\frac{1}{2} \partial_t \Xi \partial_t \Xi
 -\frac{1}{2} \partial_i \Xi \partial_i \Xi
 -B \frac{\left(\partial_i \Phi^* \partial_i \Phi\right)^2}{\Xi^6}
 - H \Xi^6 - \lambda^2 \left(\Phi^* \Phi\right)^3 \ .
\end{equation}
The additional term breaks explicitly the shift symmetry, and would introduce an explicit dependence on the coordinates in the effective action of the meta-fluid. 

The equations of motion are given by
\begin{align}
 \partial_t^2 \Phi
 + A\,  \partial_i^2 \Phi
 - 2B\, \partial_i \left(\frac{\partial_i \Phi}{\Xi^6} \partial_j \Phi^* \partial_j \Phi \right) + 3 \lambda^2 \Phi^{*2} \Phi^3
 &= 0 \ , \\
 \partial_t^2 \Xi
 - \partial_i^2 \Xi
 - \frac{6}{\Xi} \left[B \frac{\left(\partial_i \Phi^* \partial_i \Phi\right)^2}{\Xi^6}
 - H \Xi^6\right]
 &= 0 \ .
\end{align}
To achieve a similar spontaneous symmetry breaking pattern as in Sections \ref{GroundStatesSection} and \ref{WISection}, we mimic the helical ansatz \eqref{ansa1}, \eqref{ansa2} where the chemical potential is implemented by a time-dependent phase in the $U(1)$-direction. Written explicitly, it provides 
\begin{align}\label{ansa1mu}
 \Phi(t,x,y) &= \rho\, e^{i( \mu t +  k x)}\ ,\\ \label{ansa2mu}
 \Xi(t,x,y) &= v\ ,
\end{align}
where the parameters $v$, $\rho$, $k$ and $\mu$ are all real and non vanishing, and assumed to be positive for simplicity. The equations of motion are 
\begin{align}\label{eom1}
  \rho ^2 \left(A k^2-\frac{2 B k^4 \rho ^2}{v^6}-3 \rho ^4 \lambda^2+\mu ^2\right) = 0\ ,\\ \label{eom2}
 B k^4 \rho ^4-H v^{12} = 0\ .
\end{align}
We keep the same relation between the coefficients of the action
\begin{equation} \label{FineTuning}
 H=\frac{A^2}{4B}\ ,
\end{equation}
so that the relation \eqref{para} remains unchanged, but there is an additional condition
\begin{equation}
    \mu^2=3 \rho^4 \lambda^2 \label{EOMImplications} \ .
\end{equation}
Therefore $\rho$ is fixed in terms of $\mu/\lambda$. The zero density limit can be taken keeping $\rho$ fixed if both $\mu$ and $\lambda$ are taken to zero at the same rate.

The chemical potential $\mu$ is seen as an external parameter that fixes the ensemble. Therefore, $\rho$, $k$ and $v$ are parameters of the solution that should be solved in terms of $A$, $B$ and $\mu$. This can alternatively be achieved by minimizing the effective potential:\footnote{The terminology ``effective" comes from the fact that at finite density, it is customary to look for ground states which minimise the effective Hamiltonian $\Tilde{H}=H- \mu Q$ where $Q$ is the $U(1)$ conserved charge. This formulation is equivalent to searching for ground states of the Hamiltonian $H$, evolving in time along the $U(1)$-direction. Our ansatz \eqref{ansa1mu}, \eqref{ansa2mu} is precisely doing so, and by considering $\mu$ as being an external parameter, $\rho$, $k$ and $v$ parametrize a static solution minimizing the effective Hamiltonian.   }
\begin{equation}
 V_{\text{eff}}= \frac{B}{v^{6}}\left( k^2\rho^2-\frac{A}{2B}v^6\right)^2+\lambda^2 \rho^6-\mu^2\rho^2 \ .
\end{equation}
In the present case the ratio $v/\rho$ is fixed by $k/\mu$, more precisely $v^6/\rho^6\sim  k^2/\mu^2$. Then, if $k\gg \mu$ we expect the results to be quite similar to the $\mu=0$ case with $v/\rho \gg 1$, in which case the gapless mode would be mostly $\tau$. On the other hand, for $\mu\gg k$ they are expected to be closer to the case $v/\rho\ll 1$, where the mode with a gap proportional to $k$ is mostly $\tau$.

\subsection{Dispersion relations}

We are now ready to perform the fluctuations around our background. 
The fluctuations are parameterized as follows:
  \begin{align}\label{flu1mu}
 \Phi(t,x,y) &= \rho\, e^{i (\mu t + k x)} \left[1 + \phi(t,x,y)\right]
 = \rho\, e^{i (\mu t + k x)} \left[1 + \sigma(t,x,y) + i \chi(t,x,y)\right]\ ,\\ \label{flu2mu}
 \Xi(t,x,y) &= v\, \left[1 + \tau(t,x,y)\right]\ .
\end{align} 
The linearized equations of motion are\footnote{We remind that here and in the following we keep using the parameters that make the expressions simplest. However, we must always recall that the relations \eqref{para} and \eqref{EOMImplications} hold.}
\begin{align}\label{eq1}
 2A (k+i\partial_x) \left[k (\sigma -3 \tau )+\partial_x \chi\right] +\partial_t^2 (\sigma -i \chi)+2 i  \mu \partial_t  (\sigma
   -i \chi )+4 \mu ^2 \sigma &=0\ ,\\ \label{eq2}
 2A (k-i\partial_x) \left[k (\sigma -3 \tau )+\partial_x \chi\right] +\partial_t^2 (\sigma +i \chi)-2 i \mu \partial_t  (\sigma
   +i \chi )+4 \mu ^2 \sigma &=0\ ,\\ \label{flu_3_mu}
  v^2   \left(-\partial_x^2-\partial_y^2 +\partial_t^2\right)\tau-12A k \rho^2  \left[k (\sigma -3 \tau )+ \partial_x \chi \right] &=0\ .
\end{align} 
Notice that the term $4 \mu ^2 \sigma$ in \eqref{eq1} and \eqref{eq2} spoils the space-modulated shift symmetry we had in the case $\mu=0=\lambda$. We therefore do not expect a trivial mode in the spectrum. 

In Fourier space, the kinetic matrix for the fluctuations is
\begin{equation}
 M = \rho^2\left(
\begin{array}{ccc}
 {\omega^2-2A k^2-4 \mu^2} &
 -{2 i (A k q_x+\omega \mu )} & 
 6A k^2 \\
 {2 i (A k q_x+\omega \mu )} &
 {\omega^2-2A q_x^2} & 
 -6 i A k q_x \\
 6A k^2 & 
 6 i A k q_x & 
 \frac{1}{2} \left(\omega^2-q_x^2-q_y^2\right) \frac{v^2 }{\rho^2} - 18A k^2 
\end{array}
\right)\ ,
\label{matFiniteDensity}
\end{equation}
where, as before, the first line corresponds to $\sigma$, the second one to $\chi$ and the third one to $\tau$. 
Its determinant is given by
\begin{align}
 &\det M=\frac{\rho^4}{2}
 \Big\{\omega^2 v^2  \left(\omega^2-8 \mu ^2\right)
   \left(\omega^2-q_x^2-q_y^2\right) \\ \nonumber
   & \qquad -2A \left[v^2 
   \left(\omega^2-q_x^2-q_y^2\right) \left(\omega^2 \left(k^2+q_x^2\right)+4
   k \omega q_x \mu -4 q_x^2 \mu ^2\right)+18 k^2 \rho^2 \omega^2
   \left(\omega^2-8 \mu ^2\right)\right]\Big\}\ .
   \label{detMatFiniteDensity}
\end{align}
Setting the momenta to zero, one gets
\begin{align}
 \det M=\frac{\rho^4\omega^2}{2} \left[\omega^2 v^2 \left(\omega^2-2A k^2-8 \mu ^2\right)
 -36A k^2 \rho^2 \left(\omega^2-8 \mu ^2\right)\right]\,,
\end{align}
whose zeros give the mass spectrum. One thus finds a gapless mode $m_1^2=0$ and two gapped modes, whose gaps are 
\begin{equation}\label{gaps}
    \begin{split}
     m_{2,3}^2 &=  A k^2 \left(18 \frac{\rho ^2}{v^2}+1\right)+4\mu^2 \mp \sqrt{\left[Ak^2 \left(18 \frac{\rho ^2}{v^2}+1\right)+4\mu^2\right]^2-288A
    k^2 \mu^2 \frac{\rho^2}{v^2}}\ .
    \end{split}
\end{equation}
The reduction of the number of massless modes compared to the zero-density case is expected due to the explicit breaking of the shift symmetry by the pair $\mu$ and $\lambda$. Intuitively, such breaking leads to one less flat direction and hence, to one fewer gapless mode.

If we take $\mu\ll k$ while keeping $\rho$ and $v$ fixed, one gets
\begin{align}\label{sero}
 m^2_2 &=\frac{144\rho^2}{18\rho^2+v^2}\mu^2+{\cal O}\left(\frac{\mu^4}{k^2}\right)\ , \\ \label{reso}
 m^2_3 &= m_\eta^2  + {\cal O}(\mu^2)\ ,
\end{align}
where we recall that $m_\eta^2$ as given in \eqref{gap3} is of ${\cal O}(k^2)$. Note that since as we already noticed, we have that $\mu/k\sim(\rho/v)^3$, the leading term in \eqref{sero} goes to zero really as $m_2^2\sim\mu^{8/3}k^{-2/3}$. In any case, the zero density limit returns the spectrum computed in Section \ref{DispRelHelicalSection} as expected. 

In the opposite limit, $k\ll \mu$, we have
\begin{align}\label{serok}
 m^2_2 &=36Ak^2\frac{\rho^2}{v^2}+{\cal O}\left(\frac{k^4}{\mu^2}\right)\ , \\ \label{resok}
 m^2_3 &= 8\mu^2  + {\cal O}(k^2)\ .
\end{align}
Again, note that taking into account the behaviour of $\rho/v$ in the limit, we have that $m_2^2\sim k^{4/3}\mu^{2/3}$, still very much suppressed with respect to $m_3^2\sim \mu^2$. 
The upshot is thus that in both limits there is a large separation between the larger and smaller gap $m_3\gg m_2$.

\begin{figure}[!ht] 
 \begin{center}
  \includegraphics[width=\textwidth]{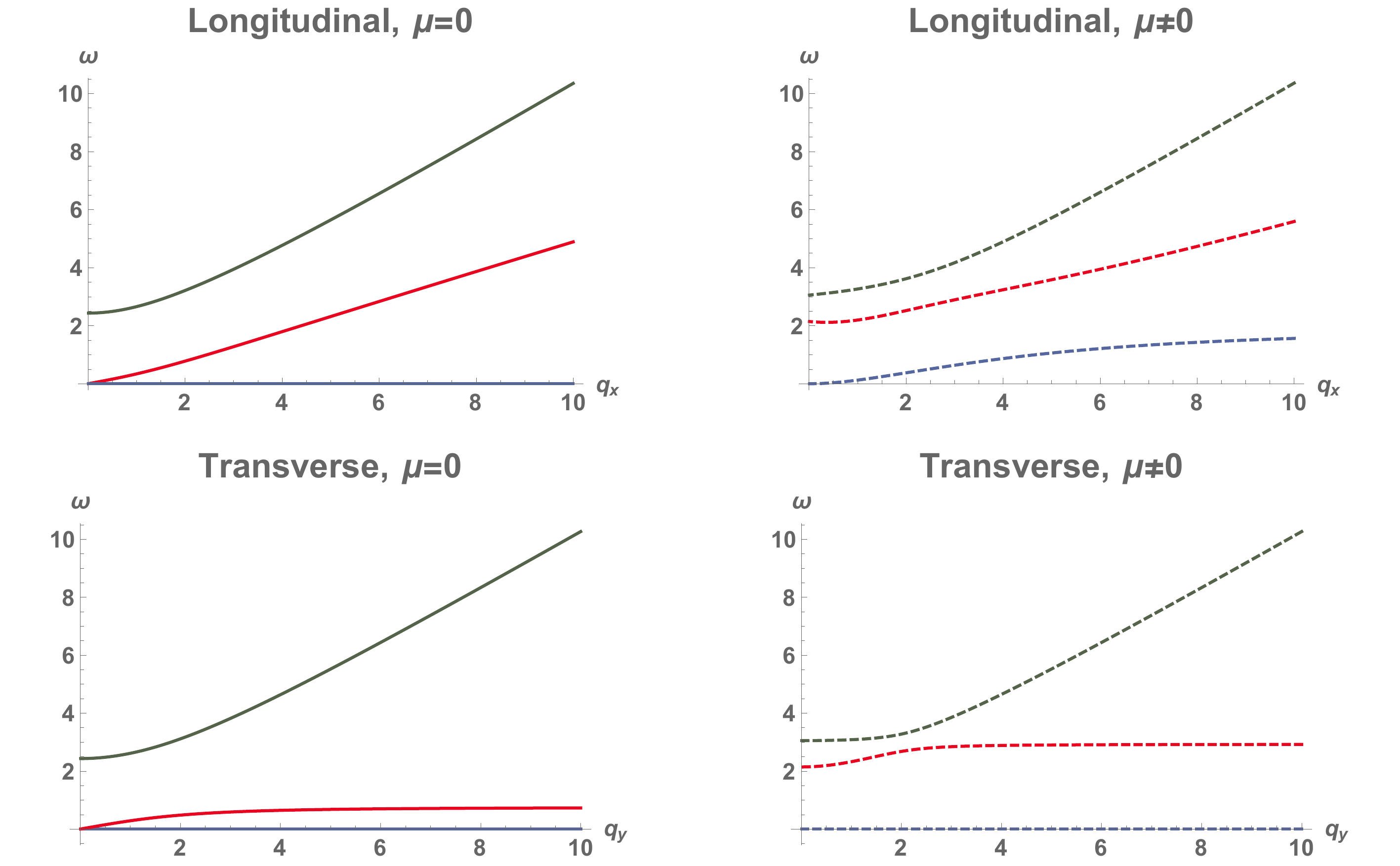}
 \end{center}
 \caption{This figure displays the dispersion relations of the three modes $\omega_1$ (blue), $\omega_2$ (red) and $\omega_3$ (green). They have been obtained by a numerical analysis of the roots of the determinants \eqref{detMat} and \eqref{detMatFiniteDensity}. The array of plots is such that each line corresponds respectively to the longitudinal direction ($q_y=0$) and the transverse direction ($q_x=0$). The columns refer to the case of zero and non-zero chemical potential -- to make it more visual, the zero chemical plots are the solid curves while the non-zero chemical ones are dashed.  All plots are done with $A=0.125$, $k=1.5$; the left column is obtained with $\mu= 0 =\lambda$ while the right column is obtained with $\mu= 1$ and $\lambda=0.5$. The VEV value $\rho$ is fixed in the $\mu\neq 0$ case by the preceding cited parameters but it is not so in the zero chemical potential case. For practicality, we took the same value for $\rho$ in both cases. Since $k>\mu$, it means that  $v>\rho$. Hence, at low momentum, the green curve is mostly shiftonic while the red curve is mostly dilatonic.}
   \label{ToyModelDispRel}
\end{figure}

\begin{figure}[!ht] 
\begin{center}
\includegraphics[scale=0.4]{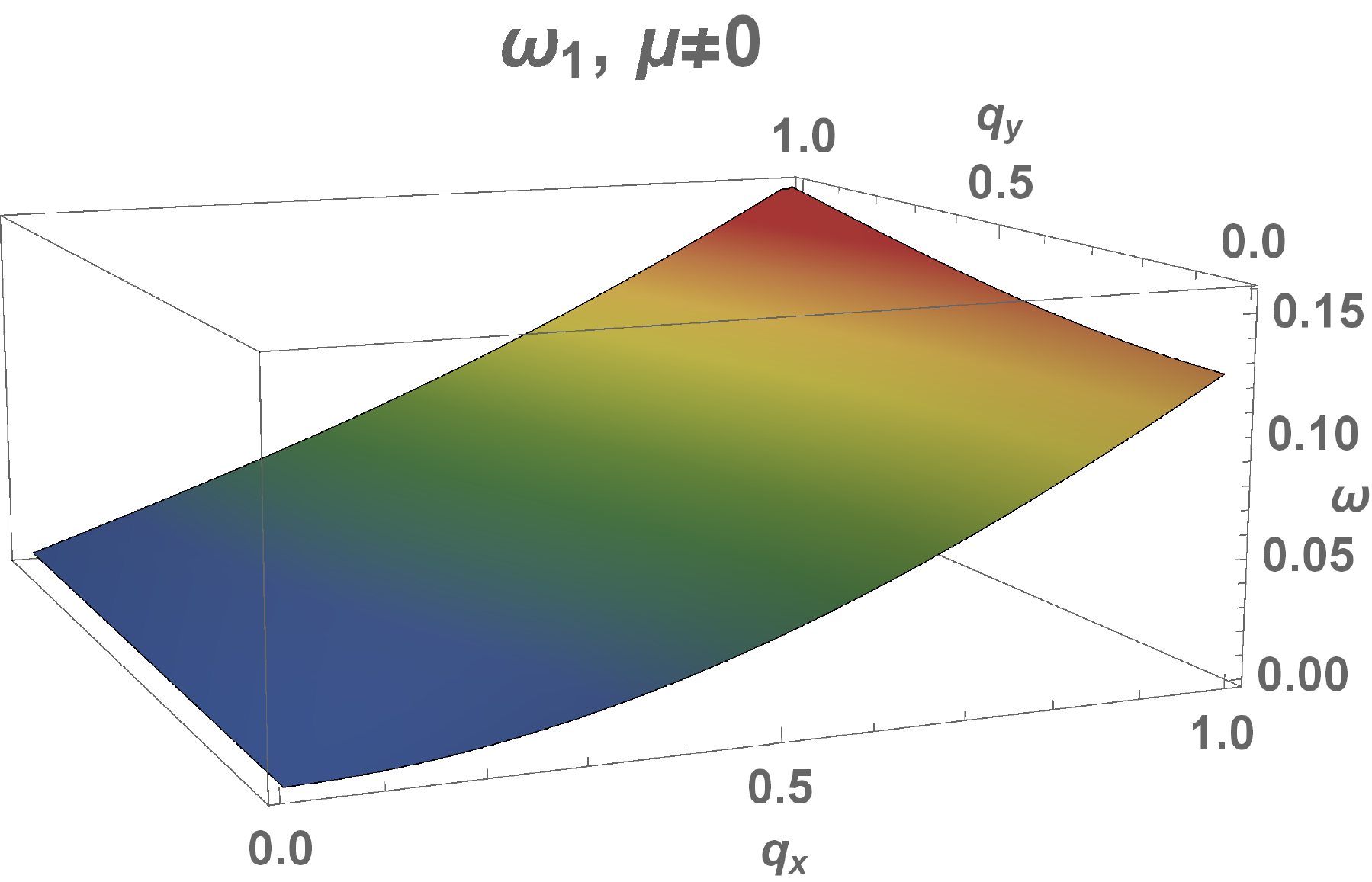} \;
\includegraphics[scale=0.4]{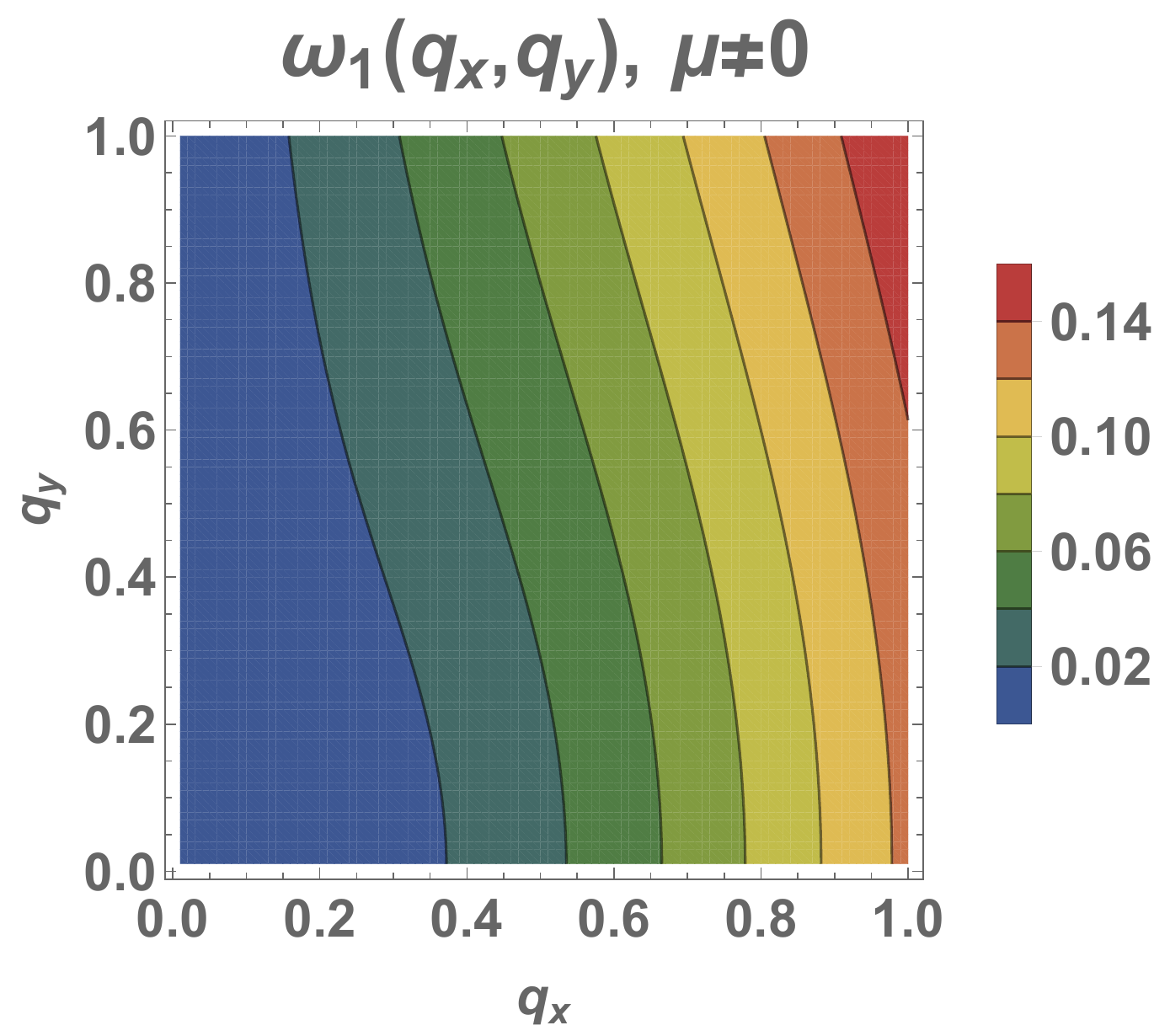} 
 \caption{This figure displays the numerical $\omega_1$ mode at low momentum, \emph{i.e.} the $U(1)$ Nambu-Goldstone mode. Both plots represent the same graph and have been obtained with $A=0.125$, $k=1.5$, $\mu= 1$ and $\lambda=0.5$. On the left, a 3D plot is provided while on the right it is a contour plot.}
   \label{ShiftonMu}
 \end{center}
\end{figure}

It is also possible to compute analytically the dispersion relation for $\omega_1$, by taking the limit $\omega, q\ll \mu, k$ in \eqref{detMatFiniteDensity}. Actually, one can solve the resulting equation by further postulating $\omega\ll q$, so that we get:
\begin{equation}
 \omega_{1}^2 
 = \frac{v^2}{36\rho^2k^2} \ q_x^2 \left(q_x^2 + q_y^2\right)
 + {\cal O}(q^5) \ . \label{MuShiftSmall}
\end{equation}
Note that this expression is valid for momenta smaller than the chemical potential. Taking the zero density limit for any fixed momentum one recovers that $\omega_1$ is the trivial mode.
At large momentum, in any non-zero $q_x$ direction, we can again solve for $\omega, k, \mu\ll q_x$ and get
\begin{equation}
 \omega_{1}^2 
 = 4 \mu^2 +  {\cal O}\left(\frac{\mu^4}{q_x^2}\right)\ , \label{MuShiftLong}
\end{equation}
while in the pure transverse direction we have the exact dispersion relation:
\begin{equation}
 \omega_{1}(0,q_y) 
 = 0\ ,  \label{MuShiftTrans}
\end{equation}
as can be seen from the fact that in this case an $\omega^2$ factorizes again in \eqref{detMatFiniteDensity}.
So, up to a correction that shifts the dispersion relation by a constant proportional to the chemical potential, $\omega_1$ should be identified with the trivial mode. 

For generic momenta, we compute the dispersion relations of the modes numerically and plot them in Figure \ref{ToyModelDispRel}.
The asymptotic dispersion relations shown in \eqref{MuShiftSmall}, \eqref{MuShiftLong} and \eqref{MuShiftTrans} match the blue curve in the numerical results of Figure \ref{ToyModelDispRel}. We further provide two three-dimensional plots of the low-momentum dispersion relations of $\omega_1$ in Figure \ref{ShiftonMu}.

\subsection{Identification of the modes}

The presence of a chemical potential does not fundamentally alter the equations of motion at very large momentum discussed in Section \ref{DispRelHelicalSection}. Therefore, we expect that the high momentum identification of the modes remains unchanged at finite density, with the trivial mode $\omega_1$ being mostly $\sigma$, the lineon $\omega_2$ or $\omega_3$ being mostly $\chi$ (depending on the value of $A$ and the direction in the momentum plane) and the relativistic mode $\omega_3$ or $\omega_2$ being mostly $\tau$.  This is confirmed by Figure \ref{ToyModelDispRel} where we have the same trends at large momentum for $\mu = 0$ and $\mu \neq 0$, the only difference being the non-zero plateau of the blue curve, which formally is produced by a subleading contribution at high momentum. 

Similarly, in the regime $k\gg \mu$ we do not expect the identification at low momenta of the modes to be significantly altered. Therefore $\omega_1$ would be mostly $\chi$ and, given that $v/\rho\gg 1$ in this regime, the lower gapped mode $\omega_2$ with gap $m_2\sim (\mu^4/k)^{1/3}$ would be mostly $\tau$ and the higher gapped mode $\omega_3$ with gap $m_3\sim k$ would be mostly $\sigma$.

Finally, when $\mu\gg k$ the most relevant terms producing the mixing of different modes is changed. Taking the matrix \eqref{matFiniteDensity} at zero momentum and $\mu\gg k$ leads to

\begin{equation}
 M_{q=0,\, \mu \gg k} = \rho^2\left(
\begin{array}{ccc}
 {\omega^2-4 \mu^2} &
 -{2 i \omega \mu } & 
 O(k^2) \\
 {2 i \omega \mu } &
 {\omega^2} & 
 0 \\
 O(k^2) & 
 0 & 
 \frac{1}{2} \omega^2 \frac{v^2 }{\rho^2} +O(k^2) 
\end{array}
\right)\ .
\label{massMatrixLargeChem}
\end{equation}
In principle both $\chi$ and $\tau$ become gapless in this limit (they are eigenvectors of $M$ with zero eigenvalue for $\omega=0$). However, taking into account the $O(k^2)$ corrections we see that $\tau$ acquires a gap proportional to $k$ while $\chi$ remains as the true gapless mode to leading order. Finally, the gapped mode with $\omega\simeq 2\sqrt{2} \mu$ is a linear combination $\sim \sigma-\frac{i}{\sqrt{2}}\chi$. Summarizing, the identification of the modes is
$$
\begin{array}{|c|c|c|c|}
\hline  &  k\gg \mu \gg q & \mu \gg k\gg q & q\gg k,\mu;\;A<1/2 \\ \hline
    \omega_1  &  \chi & \chi & \sigma\\
    \omega_2  &  \tau & \tau & \chi\\
    \omega_3  &  \sigma & \sigma-\frac{i}{\sqrt{2}}\chi & \tau\\ \hline
\end{array}
$$

\section{Removing the degeneracy}\label{sec:G}

The Mexican hat model we have studied in the previous sections has a large emergent symmetry that results in the presence of trivial modes in the spectrum. We can remove partially the emergent symmetry and generate non-trivial dispersion relations for all the modes by introducing additional terms to the action, while at the same time keeping the same symmetry breaking pattern. At fourth order in spatial derivatives and fields, there are two possible extensions\footnote{We can also have additional higher derivative terms for the real scalar $\Xi$, but since the background value of $\Xi$ is constant we are not interested in those.}
\begin{equation}
    \Delta {\cal L}= G\, \Xi^{-6} \partial_i \Phi^* \partial_i \Phi^*\partial_j \Phi \partial_j \Phi+F\, \Xi^{-6} \Phi^*\Phi\partial_ i \partial_j \Phi^* \partial_i \partial_j \Phi\ .
\end{equation}
However they do not produce qualitatively different results. For simplicity we will set $F=0$ in the following. We will then study the extended model

\begin{equation}
\begin{aligned} {\cal L} = &
 \partial_t \Phi^* \partial_t \Phi
 + A \partial_i \Phi^* \partial_i \Phi
 +\frac{1}{2} \partial_t \Xi \partial_t \Xi
 -\frac{1}{2} \partial_i \Xi \partial_i \Xi \\
 & +\frac{1}{\Xi^6}\left[- B \left(\partial_i \Phi^* \partial_i \Phi\right)^2 + G\, \partial_i \Phi^* \partial_i \Phi^*\partial_j \Phi \partial_j \Phi \right]  - H \Xi^6 \ ,
\end{aligned}
\end{equation}
where the $G$-term is the additional part. 

In general the new term will change the energy of the solutions. If we want to ensure that the helical superfluid background  \eqref{ansa1}, \eqref{ansa2} is a minimal energy solution we have to modify the relation between the coefficients to
\begin{equation}\label{eom12F=0}
 H=\frac{A^2}{4(B-G)}\ .
\end{equation}
With this choice the dimensionless combination 
\begin{equation}\label{paraF=0}
 \xi = \frac{k^2 \rho ^2}{v^6}\ ,
\end{equation}
remains fixed as a function of the coefficients of the action through the relation
\begin{equation}\label{FixingXi}
 A= 2 (B-G)\xi \ . 
\end{equation}
Equation \eqref{paraF=0} leaves therefore a moduli space with two flat directions since the static energy on-shell is identically zero. 

Using the same basis of fluctuations for the helical superfluid \eqref{flu1}--\eqref{flu2}, the new term introduces a contribution to the quadratic action of the form
\begin{equation}\label{eq:LGhs}
    {\cal L}_G=-4G \xi  (\partial_y \sigma)^2.
\end{equation}
This breaks partially the symmetry characterized by the transformation $\beta$ in \eqref{eq:fractonsym}. With the new term, $\beta$ is restricted to be a function at most linear in $y$, but yet arbitrary in $x$. So the emergent symmetry with nonzero $G$ is
\begin{equation}\label{eq:fractonsymwithG}
    \delta\chi=\alpha(y)+\beta(x) + \epsilon(x) y, \ \ \delta\sigma=-\frac{1}{k}\left[\beta'(x)+ \epsilon'(x) y\right]+3\delta+3\gamma_i x^i, \ \ \delta\tau(x,y)=\delta+\gamma_i x^i.
\end{equation}
Following the same reasoning we did previously, we expect $\chi$ and $\sigma$ to be both lineons, with $\chi$ moving along the $x$ direction and $\sigma$ along the $y$ direction.

For the meta-fluid the ansatz \eqref{eq:metaansatz} introduces a term in the action for fluctuations
\begin{equation}\label{eq:LGmf}
    {\cal L}_G\simeq \frac{2G(|b|^2)^2}{v^6}u_{ij}u_{ij}=\frac{4G(|b|^2)^2}{v^6}\partial_i u_{j}\partial_iu_{j}\ ,
\end{equation}
where $u_{ij}=\partial_i u_j+\partial_j u_i-\delta_{ij}\partial_k u_k$ is the shear strain, and the second equality is obtained up to total derivatives. Note that, in contrast to the helical superfluid,  we do not have to change the relation of $H$ with the other coefficients since this term does not give a contribution to the energy density of the background. Furthermore, the meta-fluid is stable for $G<0$, while the helical fluid is stable for $G>0$, since this gives the right sign to the kinetic terms in \eqref{eq:LGhs} and \eqref{eq:LGmf}. So, with the new term, only one of the two states would be realized depending on the values of the coefficients we choose to extend the model.

In the meta-fluid the new term introduces a shear modulus ${\cal G}=-4G|b|^2/v^6$, that removes most of the symmetries in \eqref{eq:fractusym} and \eqref{eq:rotshear}, leaving just the symmetries for massless fields.

Integrating out the massive dilaton as before will remove the zero-momentum bulk modulus, but the higher derivative terms only affect the dispersion relation at higher order in momentum. Then, the effective low-energy theory is almost the same as ordinary elasticity, the dispersion relation for the fluctuations $u_i$ is at lowest order in momentum
\begin{equation}
    \omega^2\simeq {\cal G}q^2\ .
\end{equation}

\subsection{Ward-Takahashi identities}

\subsubsection*{Helical superfluid}

For the helical superfluid the $U(1)$ Ward-Takahashi identity \eqref{con_cor} does not change when $G$ is introduced, so the $U(1)$ Nambu-Goldstone mode has the same dispersion relation at high momentum. The real shift symmetry Ward-Takahashi identity becomes
\begin{align}
\partial_t^2 \sigma-4\xi  G \partial_y^2 \sigma+2Ak \left[k(\sigma-3\tau)+\partial_x \chi\right] &=0\ .
\end{align}
The dilatation Ward-Takahashi identity also acquires a new contribution
\begin{equation}
    v^2(\partial_i^2\tau-\partial_t^2\tau)=2\rho^2\left( \partial_t^2\sigma-4 \xi G \partial_y^2 \sigma\right)+8 k \rho^2 A \left(k(3\tau-\sigma)-\partial_x\chi \right)\ .
\end{equation}
Since $\xi$ is a fixed quantity, in the high momentum limit $k\to 0$, the dispersion relation of the shifton is modified to
\begin{equation}\label{eq:WIlineony}
    \omega^2_\sigma \simeq 4\xi G q_y^2=\frac{2A G}{B-G}q_y^2\ .
\end{equation}
On the other hand the dilaton keeps a relativistic dispersion relation in this limit. This confirms our analysis of the emergent symmetries where we predicted that $\sigma$ would behave as a lineon moving along the $y$ direction.

\subsubsection*{Meta-fluid}

When $G$ is introduced, the dilatation Ward-Takahashi identity for the meta-fluid does not change, but there is a new term in the complex shift Ward-Takahashi identity \begin{equation}
    v(\partial_t^2 u_i-A\partial_i \partial_k u^k-{\cal G} \partial_k^2 u_i)+6 A \partial_i\tau=0.
\end{equation}
At high momentum this gives two modes with dispersion relations
\begin{equation}
    \omega^2 \simeq {\cal G}q^2,\qquad \omega^2\simeq (A+{\cal G}) q^2,
\end{equation}
with the first mode corresponding to the transverse and the second to the longitudinal components of $u_i$.

\subsection{Dispersion relations for the helical superfluid}

We obtain the following equations of motion at linear order for the fluctuations:
\begin{align}\label{primExt}
 2 A (k + i \partial_x) \left[k (\sigma -3 \tau )+ \partial_x \chi\right] + \partial_t^2 (\sigma -i \chi ) - 4 G\xi \partial_y^2 \sigma &=0\ ,\\ \label{seguExt}
 2 A (k - i \partial_x) \left[k (\sigma -3 \tau )+ \partial_x \chi\right] + \partial_t^2 (\sigma +i \chi ) - 4G\xi \partial_y^2 \sigma &=0\ ,\\ \label{tercExt}
 12 A k \rho^2 \left[k (\sigma -3 \tau )+ \partial_x\chi \right]-v^2 \left(-\partial_x^2 - \partial_y^2 + \partial_t^2\right) \tau &= 0\ .
\end{align}
Sending the parameter $G$ to zero (keeping the parameter $k$, $\rho$ and $v$ fixed) permits to recover the vacuum as well as the equations of motion of the $G=0$ model \eqref{lagra}. Hence, in this specific limit, we expect to recover smoothly the original spectrum.

Going to Fourier space, we obtain a homogeneous algebraic system for the equations of motion driven by the kinetic matrix:
\begin{equation}
 M = \rho^2 \left(
\begin{array}{ccc}
 \omega^2-2 A k^2 - 4 G\xi q_y^2 & -2 i A k q_x  & 6 A k^2 \\
 2 i A k q_x & \omega^2-2 A q_x^2 & -6 i A k q_x \\
 6 A k^2 & 6 i A k q_x & \frac{1}{2}\left(\omega^2-q_x^2-q_y^2\right) \frac{v^2}{\rho^2} -18 A k^2 
\end{array}
\right)
\label{MF=0}
\end{equation}
The determinant evaluates to
\begin{align}
    \det (M)=\det (M)_{G=0} -2G\xi \rho^4q_y^2\left[v^2 (\omega^2-q_x^2-q_y^2)(\omega^2-2Aq_x^2)-36A{k^2\rho^2}\omega^2\right]\ ,
\end{align}
where $\det(M)_{G=0}$ is given by \eqref{detMat}.

If we specifically look for a trivial root of the determinant, we do not find one:  
\begin{equation}
\text{det}(M)|_{\omega=0} = -4 A G \xi \rho ^4 v^2 q_x^2 q_y^2 \left(q_x^2+q_y^2\right) \ .
\end{equation}
This immediately tells us that there is no longer a trivial mode. This is consistent with the analysis we made based on the emergent shift symmetries. 

The spectrum features two gapless modes and one gapped mode
\begin{align}
 m_1^2 &=0\ , \\
 m_2^2 &= 0\ ,\\ \label{gap3F=0}
 m_3^2 & = 2 A k^2 \left(1+18 \frac{\rho^2}{v^2}\right)\ .
\end{align}
Notice that the difference with the case $G=0$ is hidden in the relation among the parameters, where the correction is given by a factor $(B-G)$ instead of simply $B$. So sending $G$ to zero smoothly provides the masses of the $G=0$ case.

Proceeding to compute the dispersion relations at small momenta, we obtain:
\begin{align}\label{zerozeroG}
 \omega_1^2 &= \frac{4 G \rho^2 v^2 (q_x^2+q_y^2)q_x^2q_y^2 }{72 k^2 \rho^4 G q_y^2 + v^8 (q_x^2+q_y^2) }+ {\cal O}(q^6)\ , \\
 \omega_2^2 &= \frac{v^2}{18\rho^2+v^2} \left[q_x^2 + \left(1+\frac{72  Gk^2 \rho^4}{v^8} \right) q_y^2\right] + {\cal O}(q^4)\ ,\\
 \omega_3^2 &= 2A k^2 \left(1+18 \frac{\rho ^2}{v^2}\right)
 +2Aq_x^2+\frac{18\rho^2}{18\rho^2+v^2} 
 \left[q_x^2 + \left(1+\frac{4  Gk^2 \rho^2}{v^6} \right) q_y^2\right]
 +O\left(q^4\right) \ .
\end{align}
We recover smoothly the $G=0$ case in the limit of zero $G$.

\begin{figure}[!ht] 
 \begin{center}
  \includegraphics[scale=0.3]{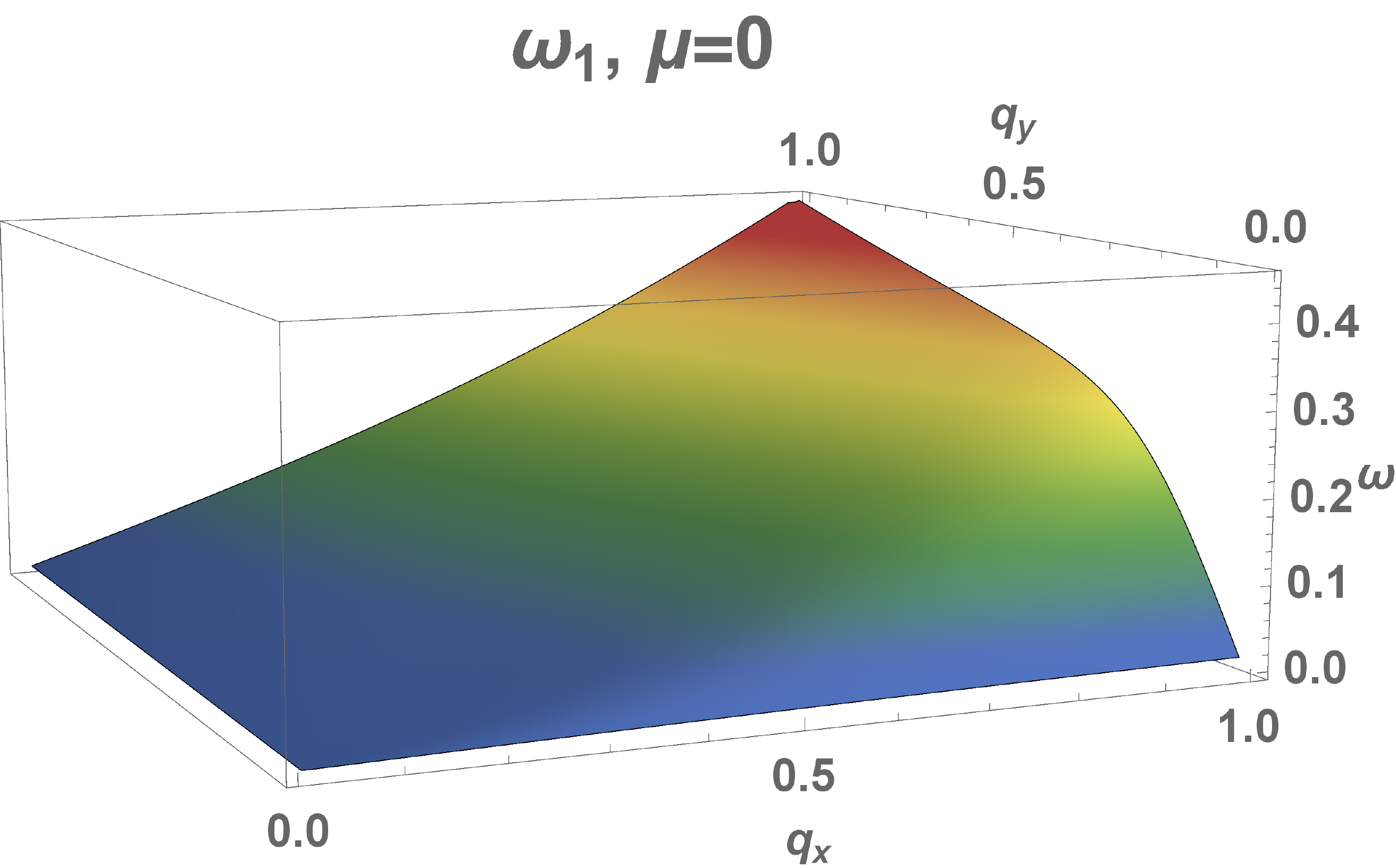}
  \includegraphics[scale=0.4]{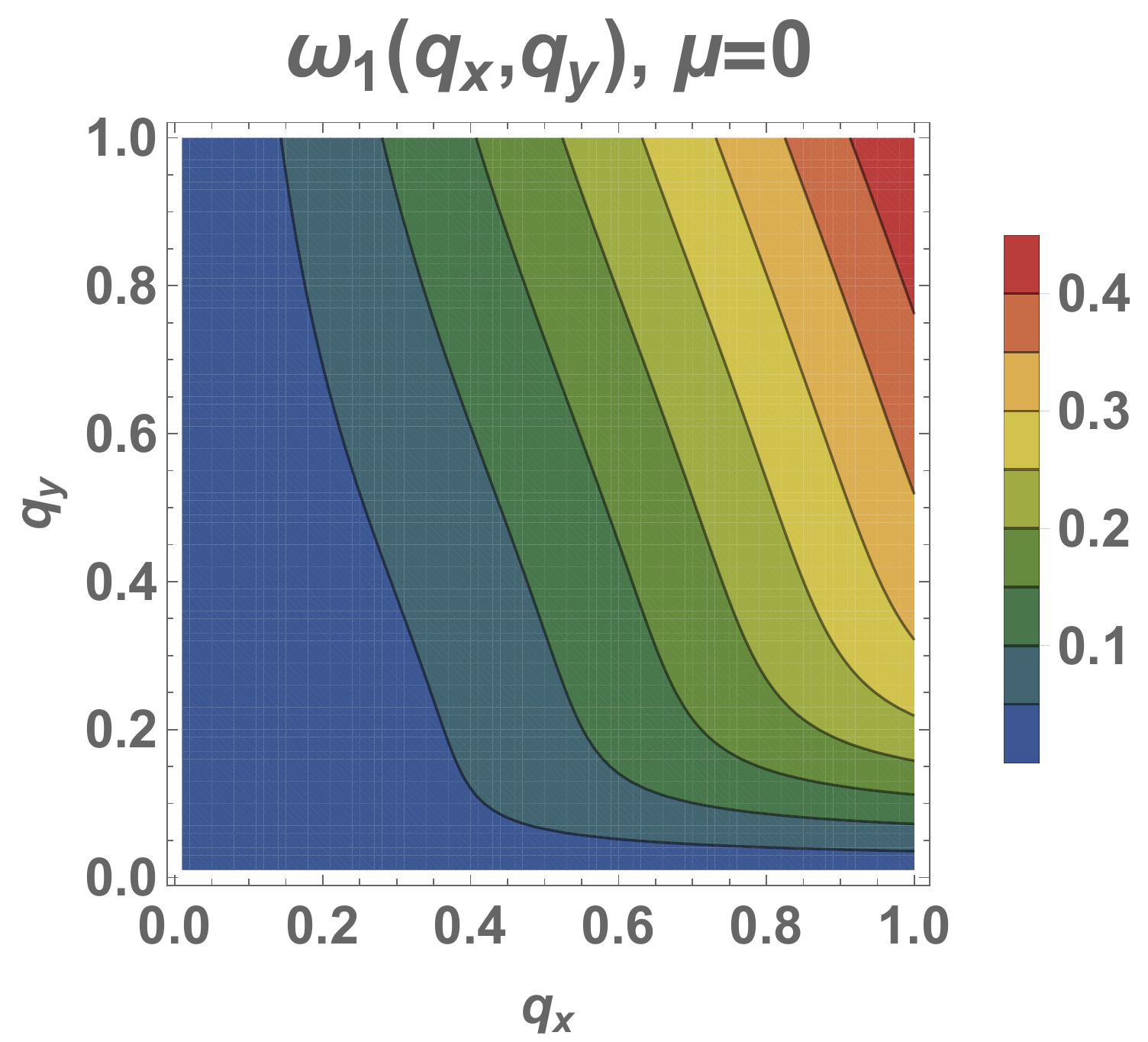}
 \end{center}
 \caption{This figure displays the numerical $\omega_1$ mode at low momentum, \emph{i.e.} the $U(1)$ Nambu-Goldstone mode. Both plots represent the same graph and have been obtained with $A=2$,  $G=1$, $k=1$, $\rho= 0.54$ and $\mu= 0 =\lambda$. On the left, a 3D plot is provided while on the right it is a contour plot. 
 }
   \label{ShiftonG}
\end{figure}

Note that the expression for the mode $\omega_1$ has an unusual non-analytic dependence with momentum but overall it goes like $\omega_1\sim q^2$, while the other gapless mode is linear $\omega_2\sim q$. The dispersion relation of $\omega_1$ is confirmed by the numerical study of Figure \ref{ShiftonG}. In fact, the plots display the trivialization of the dispersion relations for $q_x=0$ and $q_y=0$, and the non trivial bump in the quadrant in between. Also, the analytic expression predicts the changes in slope we observed in the 3D plot. Indeed, according to \eqref{zerozeroG}, the starting slope of the dispersion relation at fixed $q_x>0$ is larger than the starting slope at fixed $q_y>0$. Hence, in order for the dispersion relations to join continuously, the fixed $q_x>0$ dispersion relation should bend downwards.

A final comment is that at large momentum we observe that \eqref{MF=0} diagonalizes. In particular, the modes $\omega_1$ and $\omega_2$ are respectively transverse and longitudinal lineons with the dispersion relations 
\begin{equation}
    \omega_1\sim 2\sqrt{G\xi}\, q_y \ , \quad \omega_2\sim \sqrt{2A}\, q_x \quad \text{when } q_x,q_y\gg k\ .
\end{equation}
This is in agreement with the analysis of Ward-Takahashi identities we did previously.

\subsubsection*{Identification of the modes}

Since the new term proportional to $G$ adds a contribution to the kinetic matrix proportional to $q_y^2$, the separation between gapless and gapped modes at low momentum is the same as for $G=0$. Indeed, the gap in \eqref{gapped} and the velocity in the $x$ direction of the gapless mode \eqref{gapless} are the same with nonzero $G$ \eqref{zerozeroG}. The nature of the high momentum modes is easily identified with the help of the Ward-Takahashi identities. So the identification of the modes is essentially the same as in section \ref{sec:idenmod}, except $\omega_1$ has a non-analytic behavior at low momentum and becomes a lineon propagating in the $y$ direction at high momentum.

\subsection{Dispersion relations for the meta-fluid}

Let us consider directly the kinetic matrix 
\begin{equation}
M=|b|^2 \left(
\begin{array}{ccc}
 \omega^2-Aq_x^2-{\cal G}q^2
 & -Aq_xq_y
 & -6i \frac{A}{v} q_x \\
 -Aq_xq_y & \omega^2-Aq_y^2-{\cal G}q^2 
   & -6i \frac{A}{v} q_y \\
 6i \frac{A}{v} q_x
 & 6i \frac{A}{v} q_y & \frac{1}{2|b|^2} \left(\omega^2- q^2\right) - 36\frac{ A }{v^2}
\end{array}
\right)\ , \label{KineticMatrixMetaDensity}
\end{equation}
where we recall that ${\cal G}=-4G|b|^2/v^6>0$. 
We notice that only the first two diagonal terms are modified compared to the $G=0$ case. Therefore, we expect that only two of the three dispersion relations will be more significantly affected by the correction, namely the lightest modes. 

The determinant of the kinetic matrix reads as follows
\begin{align}\label{detmetafluidwithG}
    \det (M)= \frac{|b|^4}{2} (\omega^2-{\cal G} q^2) \Big[ (\omega^2-q^2)(\omega^2-Aq^2-{\cal G}q^2)-m_\tau^2(\omega^2-{\cal G}q^2)\Big]\ .
\end{align}
It rightly reduces to \eqref{detmetafluid} when $G=0$.
From this expression one can immediately see that what was formerly the immobile fracton, acquires isotropic and linear dispersion relations which are valid for any momenta, and are entirely controlled by $G$. One can further find the exact analytical expression for the other two modes, which will depend non-trivially both on $G$ and $m_\tau$.
At low-momentum $\omega, q\ll m_\tau$, one can see that the condition $\det (M)=0$ gets an additional factor of $(\omega^2-{\cal G}q^2)$, giving the two gapless modes expected from the low-energy effective theory.

In more detail, at low momentum we have the expansions
\begin{align}
 \omega_1^2 & = {\cal G} q^2 \ ,\label{trivialMetaDensity}\\
 \omega_2^2 &= {\cal G}q^2+A(1-{\cal G})\frac{q^4}{m_\tau^2}  +{\cal O}\left(q^6\right)\ ,\label{nonTrivialMetaDensity}\\
 \omega_3^2 &= m_\tau^2
 + (1+A) q^2 + {\cal O}\left(q^4\right)\ .  \label{massiveMetaDensity}
\end{align}
We recover the expected results from the effective analysis as well as the idea that two of the three modes are more substantially affected by $G$.  Looking at \eqref{trivialMetaDensity}, \eqref{nonTrivialMetaDensity} and \eqref{massiveMetaDensity} we get back the dispersion relation of the original model when we send $G$ to zero. 

At large momentum we can drop the last term in \eqref{detmetafluidwithG}, so that the determinant completely factorizes and the modes will behave as 
\begin{align}
 \omega_1  = \sqrt{{\cal G}}|q| \ , \qquad \omega_2 \simeq \sqrt{(A+{\cal G})}|q|\ , \qquad \omega_3 \simeq |q|\ .
\end{align}
When $G \rightarrow 0$, we recover the original large momentum behavior. 

According to our previous analysis, at low momentum we can identify $\omega_1$ with the transverse component of the displacement $u_i$, while $\omega_2$ corresponds to the longitudinal part. The gapped mode is mostly the dilaton. At high momentum the Ward-Takahashi identities keep the same identification for the modes as for low momentum.

\subsection{Extended model at finite density}

We generalize the helical superfluid solutions to finite density, as already done in Section \ref{OrignModelFiniteDensity} for $G=0$. In order to stabilize the ground states, we add a shift symmetry breaking $\lambda$-term in the Lagrangian,
\begin{equation}
\begin{aligned}
 {\cal L} = &
 \partial_t \Phi^* \partial_t \Phi
 + A \partial_i \Phi^* \partial_i \Phi
 +\frac{1}{2} \partial_t \Xi \partial_t \Xi
 -\frac{1}{2} \partial_i \Xi \partial_i \Xi \\
 & +\frac{1}{\Xi^6}\left[- B \left(\partial_i \Phi^* \partial_i \Phi\right)^2 + G\, \partial_i \Phi^* \partial_i \Phi^*\partial_j \Phi \partial_j \Phi \right]  - H \Xi^6 
 - \lambda^2 \left(\Phi^* \Phi\right)^3 \ .
 \label{lagraF=0}
\end{aligned}
\end{equation}
Given the condition \eqref{eom12F=0}, the plane wave ansatz \eqref{ansa1mu}--\eqref{ansa2mu} is a solution to the equations of motion minimizing the effective potential provided 
\begin{equation}\label{VvrhoF=0}
  v^6 = \frac{2(B-G)}{A} k^2 \rho^2 \ , \qquad  
  \rho^2 = \left|\frac{\mu}{\sqrt{3}\, \lambda}\right| \ .
\end{equation}
Setting $G$ to zero, we recover the background solution of the finite density $G=0$ model.

\begin{figure}[!ht] 
 \begin{center}
  \includegraphics[scale=0.3]{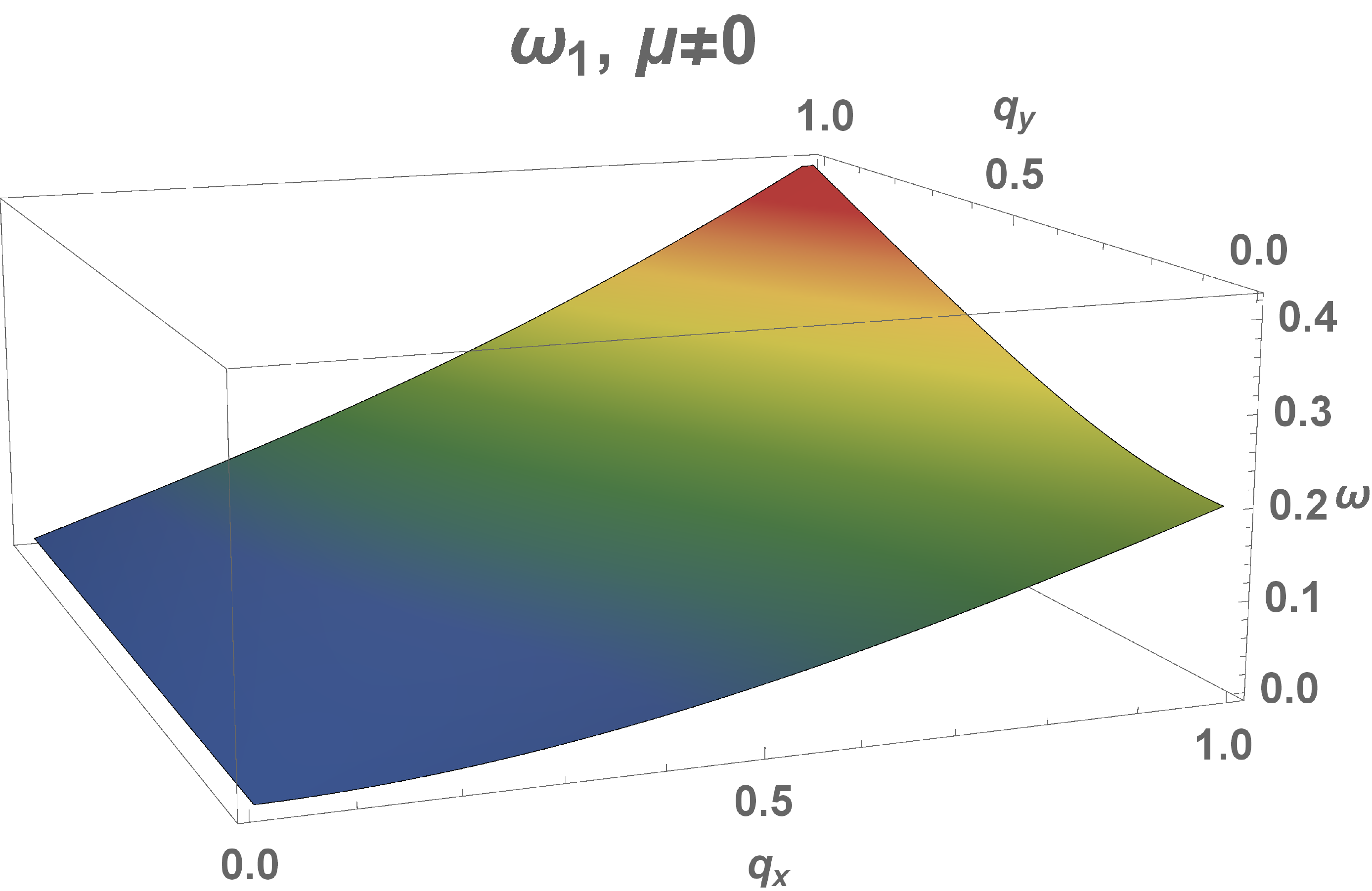}
  \includegraphics[scale=0.4]{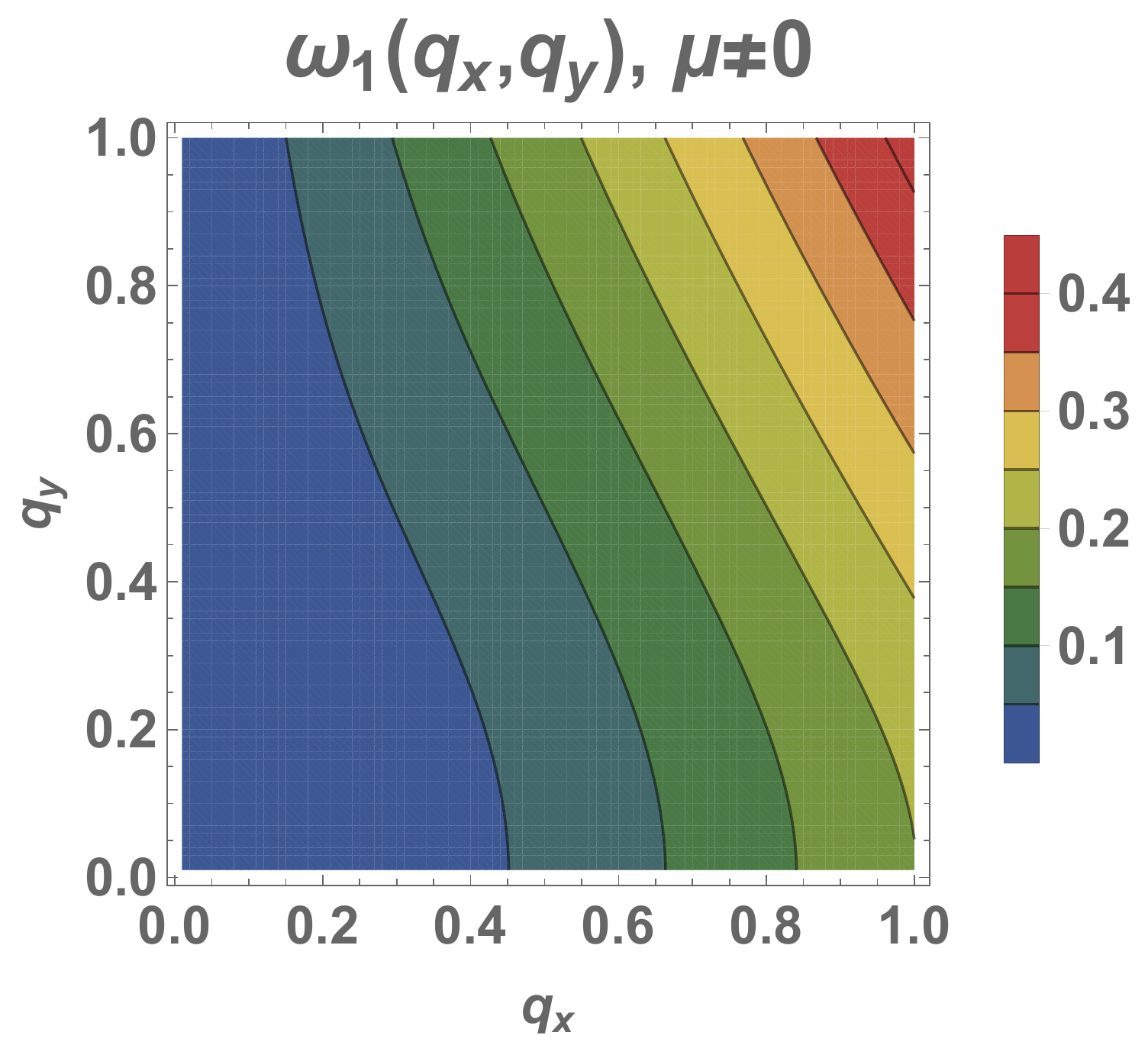}
 \end{center}
 \caption{This figure displays the numerical $\omega_1$ mode at low momentum, \emph{i.e.} the $U(1)$ Nambu-Goldstone mode. Both plots represent the same graph and have been obtained with $A=2$, $G=1$, $k=1$, $\mu= 0.5$ and $\lambda=1$. On the left, a 3D plot is provided while on the right it is a contour plot. 
 }
   \label{ShiftonGMu}
\end{figure}

The linearized equations of motion are
\begin{align}\label{eq1Ext}
 2A (k+i\partial_x) \left[k (\sigma -3 \tau )+\partial_x \chi\right] +\partial_t^2 (\sigma -i \chi)- 4 G\xi \partial_y^2 \sigma  +2 i  \mu \partial_t  (\sigma
   -i \chi )+4 \mu ^2 \sigma &=0\ ,\\ \label{eq2Ext}
 2A (k-i\partial_x) \left[k (\sigma -3 \tau )+\partial_x \chi\right] +\partial_t^2 (\sigma +i \chi)- 4 G\xi \partial_y^2 \sigma-2 i \mu \partial_t  (\sigma
   +i \chi )+4 \mu ^2 \sigma &=0\ ,\\ \label{flu_3_muExt}
  v^2   \left(-\partial_x^2-\partial_y^2 +\partial_t^2\right)\tau-12A k \rho^2  \left[k (\sigma -3 \tau )+ \partial_x \chi \right] &=0\ .
\end{align}
The kinetic matrix associated to the equations of motion is
\begin{equation}
 M = \rho^2\left(
\begin{array}{ccc}
 \omega^2-2A k^2-4 \mu^2 - 4 G\xi q_y^2 &
 -2 i (A  k q_x+\omega \mu ) & 
 6A  k^2 \\
 2 i (A  k q_x+\omega \mu ) &
 \omega^2-2A q_x^2 & 
 -6iA k q_x \\
 6Ak^2 & 
 6 i A k q_x & 
 \frac{1}{2} \left(\omega^2-q_x^2-q_y^2\right) \frac{v^2 }{\rho^2} - 18Ak^2 
\end{array}
\right)\ .
\end{equation}
\begin{figure}[!ht] 
 \begin{center}
  \includegraphics[width=\textwidth]{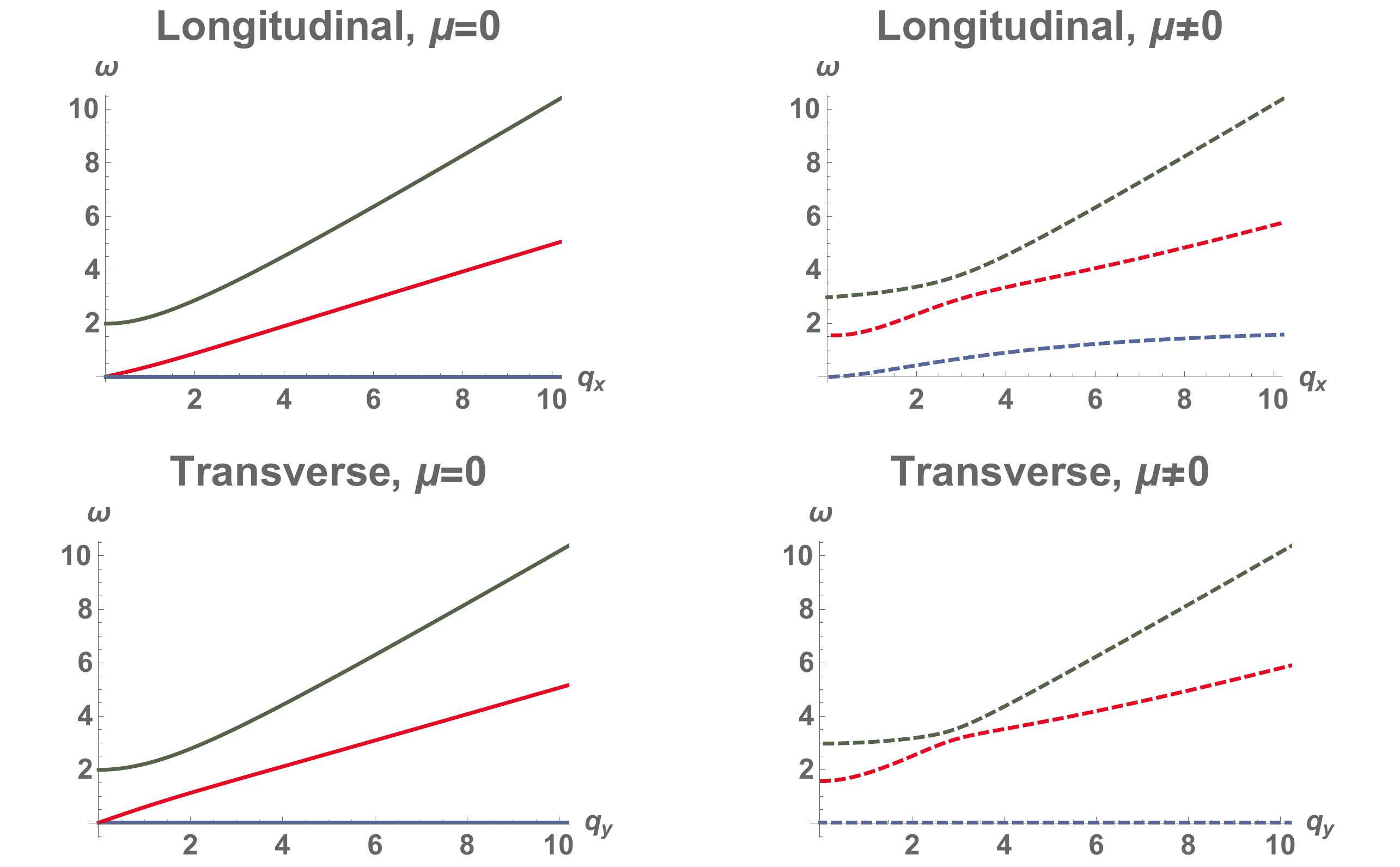}
 \end{center}
 \caption{This figure displays the dispersion relations of the three modes (each mode has its own color, as in Figure~\ref{ToyModelDispRel}). The array of plots is such that each lines corresponds respectively to the longitudinal direction ($q_y=0$) and the transverse direction ($q_x=0$). The columns refer to the case of zero and non-zero chemical potential -- to make it more visual, the zero chemical plots are the solid curves while the non-zero chemical ones are dashed.  All plots are done with $A=0.125$, $G=0.25$, $k=1.5$; the left column is obtained with $\mu= 0 =\lambda$ while the right column is obtained with $\mu= 1$ and $\lambda=1$. The VEV value $\rho$ is fixed in the $\mu\neq 0$ case by the preceding cited parameters but it is not so in the zero chemical potential case. We took the same value for $\rho$ in both cases for practical reasons. We have that $v>\rho$, hence, at low momentum, the green curve is mostly shiftonic while the red curve is mostly dilatonic. Notice that since $\mu<k$, the identification of the modes for the finite density case matches the one with zero chemical potential.}
   \label{ExtendedModelDispRel}
\end{figure}
\noindent
Since $G$ only contributes by terms proportional to the momentum, there are no qualitative differences in the gaps, it is enough to replace $B\to B-G$ in the expressions found in Section \ref{OrignModelFiniteDensity}. The high momentum behavior will once more be the same as for zero density. For low and intermediate momenta, we resort to numerics, our results are plotted in Figure \ref{ExtendedModelDispRel} and in Figure \ref{ShiftonGMu}. Comparing with Figure \ref{ShiftonG}, we observe that the $\omega_1$ mode is lifted at $q_y=0$ when $\mu\neq 0$, as also happened at $G\neq 0$. On the other hand, comparing Figure \ref{ShiftonMu} and Figure \ref{ShiftonGMu}, the effect of $G$ is to introduce a change in the slope of the dispersion relation in the $q_x$ direction. The identification of the modes will be the same as that made at $G=0$ in Section  \ref{OrignModelFiniteDensity}.

\section{Counting the Nambu-Goldstone modes}\label{sec:count}

The counting of Nambu-Goldstone modes for internal symmetries has been systematically studied through the years, also in relation to their dispersion characteristics \cite{Nielsen:1975hm,Leutwyler:1993gf,Schafer:2001bq,Watanabe:2011dk,Nambu:2004yia,Watanabe:2012hr,Kapustin:2012cr,Hidaka:2012ym,Watanabe:2014fva,Takahashi:2014vua}, and -in some cases- encompassing translational symmetries too \cite{Watanabe:2011ec} (see \cite{Watanabe:2019xul} for a review).
A comparison of the specific results found above with the general knowledge on Nambu-Goldstone counting is interesting because non-trivial. To this purpose, we recapitulate in Table \ref{tab} the Nambu-Goldstone modes found explicitly from the study of the helical fluid and meta-fluid fluctuation Lagrangians, as well as their dispersion and analyticity properties. In particular, we stress that we found in general two gapless and a gapped mode. 
\begin{table}[h!]
    \centering
    \begin{tabular}{c|ccc}
        \text{vacuum} & $\omega_1$ & $\omega_2$ & $\omega_3$ \\
        \hline
        \text{Helical} $(G=0)$ &$0$&$q_\star$& \text{gapped}  \\
        \text{Helical} $(G\neq0)$&$q^2_\star$&$q_\star$& \text{gapped}\\
        \hline
        \text{Meta-fluid} $(G=0)$ &0&$q^2$& \text{gapped}\\
        \text{Meta-fluid} $(G\neq0)$&$q_\star$&$q_\star$& \text{gapped}
    \end{tabular}
    \caption{Dispersion and analyticity properties of the Nambu-Goldstone modes as found from the low-energy study of the fluctuation Lagrangian. The $\star$ sub-index indicates non-analyticity.}
    \label{tab}
\end{table}

For internal symmetries, the number of Nambu-Goldstone modes $n_{\text{NG}}$ is generically bounded by the number of spontaneously broken symmetries $n_{\text{BS}}$. If there are no terms with single time derivatives in the effective action, however, all the Nambu-Goldstone modes are of type A in the classification of \cite{Watanabe:2012hr,Watanabe:2014fva} and we have $n_{\text{NG}}=n_{\text{BS}}$. A priori, that would be the case for the effective actions we found (we set $\mu=0$ for the moment). However, we are not dealing with internal symmetries only, hence the notion of $n_{\text{BS}}$ has to be qualified, as we will do shortly. 

An alternative classification, perhaps more pertinent to our situation, is provided by counting theorems which split the total number of Nambu-Goldstone modes $n_{\text{NG}}$ according to specific dispersion properties. Defining as type I/type II the modes with an odd/even dispersion relation, respectively, \cite{Nielsen:1975hm} established that
\begin{equation}\label{I_II}
    n_I+2 n_{II}\geq n_{\text{BS}}\ .
\end{equation}

Turning to broken spacetime symmetries, there are no general counting rules, yet it is known that the number of independent modes can be reduced \cite{Volkov:1973vd,Ivanov:1975zq,Low:2001bw,McArthur:2010zm,Watanabe:2013iia,Nicolis:2013sga,Brauner:2014aha}. In essence, if $Q_a$ denotes the generators of broken symmetries, ${P}_i$ denotes the unbroken translations and $\left\langle\Phi({\bm x})\right\rangle$ denotes the expectation value of the order parameter, then the following set of identities allows to reduce the number of independent fields 
\begin{equation}
    [P_i,Q_a]\left\langle\Phi({\bm x})\right\rangle=c_{iab}\, Q_b\left\langle\Phi({\bm x})\right\rangle\ ,
\end{equation}
where $c_{iab}$ indicates the relevant structure constant of the symmetry algebra.
This introduces a constraint such that would-be Nambu-Goldstone bosons appearing on each side of the identity are not independent in general.

Let us discuss how the explicit results found in the previous sections relate to the general counting rules.
To the counting purposes, we have to consider the following symmetries:
\begin{itemize}
    \item Spacetime: Translations $P_1$, $P_2$, rotations $R$ and dilatations $D$; 
    \item Internal: $U(1)$ transformations (phase rotations) $Q$, complex (\emph{i.e.} real + imaginary) shifts $S_R$, $S_I$. 
\end{itemize}
Let us report the symmetry content of the two kinds of ground states separately.\\

\noindent
{\bfseries Helical superfluid}: the unbroken symmetries are two translations
\begin{equation}
    P_1-kQ\ ,\qquad P_2\ ,
\end{equation}
which leaves in principle five broken symmetries $n_{\text{BS}}=5$.
However, the commutation relation of unbroken translations with the broken generators result in additional conditions 
\begin{equation}
\begin{split}
    &[P_1-k Q,D]\propto P_1,\qquad  [P_2,R]\propto P_1,\\
    &[P_1-kQ,S_R]\propto S_I,\qquad  [P_1-kQ,S_I]\propto S_R\ .
    \end{split}
\end{equation}
This would imply that rotations, dilatations and broken translations are described by a single mode, and there would be a single mode associated to both real and imaginary shifts (indeed, we had already commented earlier on about this). Effectively we would be left with a number of independent broken symmetries $n_{\text{BS}}^{(\text{in})}=2$, where the up index stands for independent. 
\\

\noindent   
{\bfseries Meta-fluid}: the unbroken symmetries are two translations and a rotation
\begin{equation}
    P_1-S_R\ ,\qquad P_2-S_I\ ,\qquad R-Q\ ,
\end{equation}
so there would be four broken symmetries in this case $n_{\text{BS}}=4$. The commutation relations of the unbroken translations with the broken symmetries would produce additional conditions
\begin{equation}
    [P_1-S_R,D]\propto P_1,\qquad [P_2-S_I,D]\propto P_2.
\end{equation}
Note that commutators with $R+Q$ result in unbroken translations. This would imply that broken translations and dilatations are described by a single mode. Effectively this reduces the number of independent broken symmetries to $n_{\text{BS}}^{(\text{in})}=2$. 

We report the values of the countings in Table \ref{tab2}, which requires some discussion.
\begin{table}[h!]
    \centering
    \begin{tabular}{c|ccc|cc|c|c}
    \text{vacuum} & $n_I$ & $n_{II}$ & $n_0$ & $n_{\text{BS}}$ & $n_{\text{BS}}^{(\text{in})}$ & $n_A=n_{\text{NG}}$  &$n_I+2n_{II}$\\  
    \hline
         \text{Helical} $(G=0)$&1&0&1&5&2&2&2 \text{ or } 3  \\
         \text{Helical} $(G\neq0)$&1&1&0&5&2&2&3 \\
         \hline
         \text{Meta-fluid} $(G=0)$&0&1&1&4&2&2&3\text{ or }4 \\
         \text{Meta-fluid} $(G\neq0)$&2&0&0&4&2&2&2 
    \end{tabular}
    \caption{Summary of the countings for the two kinds of vacuum. With $n_0$ we denote the number of trivial modes.}
    \label{tab2}
\end{table}
For $G=0$ in the helical superfluid, there is a trivial gapless mode and a type I mode. The trivial mode could be counted as type I or II.
Still in the helical case, when $G\neq 0$, we have almost the same type of modes except that the trivial mode has a non-analytic dispersion relation, but we will still consider it as type II, since at low momenta $\omega\sim O(q^2)$. In so doing, the counting rule \eqref{I_II} is satisfied if we use $n_{\text{BS}}^{(\text{in})}$ as the number of broken symmetries.

Turning to the meta-fluid case, for $G=0$ there is an analytic/type II mode and the trivial mode, which could be counted either as type I or II. On the other hand, if $G\neq 0$, the two gapless modes are type I/non-analytic. Also in this case we observe that the counting rule \eqref{I_II} is satisfied with $n_{\text{BS}}^{(\text{in})}$ as the number of broken symmetries.

Finally, switching on a chemical potential in the helical superfluid gaps one of the two massless modes. This could be seen as a consequence of introducing a mixing through a single time-derivative term in the effective Lagrangian, so effectively we would be left with a single type B Nambu-Goldstone mode in the classification of \cite{Watanabe:2012hr,Watanabe:2014fva}, which turns out to be also type II if we just count the power of the momentum and ignore the breaking of rotational invariance.  
Note that in going to finite density we introduced an additional coupling that breaks the shift symmetry, so that in this case the remaining gapped mode is a pseudo-Goldstone mode and not a true gapped Namubu-Goldstone.
A final comment is the following. A finite chemical potential can be implemented as a linear time dependence in the phase of the charged field. In so doing, however, time-translations would be broken by our choice of ensemble and not, strictly speaking, by a dynamical feature described at the level of the Lagrangian. In contrast, the space modulations that we consider in the helical case, although being formally similar in some aspects, are determined by the gradient Mexican hat potential. Some further comments on this are given in relation to ghost condensation in the next section.

\section{Discussion}\label{sec:discuss}

The concomitant breaking of dilatations and spatial translations constitutes the main focus of the present study, which adopts the framework of effective field theory. Together with dilatations and translations, an internal Abelian symmetry is broken, too. This symmetry serves two purposes, one physical and another technical. The former consists in modeling a conserved current, thus providing the possibility of considering finite density circumstances; the latter consists in the fact that the breaking of translations and a $U(1)$ symmetry to their diagonal subgroup allows for homogeneous symmetry breaking, in the spirit of Q-lattice models \cite{Donos_2014}. 

We have utilized a simple non-relativistic field theoretical setup which allows one to characterize the low-energy modes and derive their effective description in different regimes. The breaking of translations is dynamically induced by a gradient Mexican hat mechanism, namely the competition among a quadratic gradient term driving towards instability then stabilized by higher terms \cite{Musso:2018wbv}. The gradient Mexican hat, when discretized, connects to lattice models with frustrated interactions \cite{casasola2021spontaneous}.

By means of a neat particular example, we clarified the generic fact that the analysis of the modes revolves about three relevant bases: the basis given by the fluctuations of the fields appearing in the Lagrangian; the basis of the fluctuations which diagonalize the lowest-order dynamics at low energy; and the basis associated to the symmetries of the model (this latter is possibly incomplete). Whenever the connection among such bases is non-trivial, we have mixing phenomena. For instance, the Nambu-Goldstone mode associated to a specific symmetry can results from different combinations of the UV or the IR modes, as a function of momenta.

We showed the presence of two degenerate classes of vacua, one associated to a plane wave configuration and possessing a helical structure (\emph{i.e.} a global phase rotation can be compensated by a suitable translation along the wave-vector of the plane wave), the other associated to complex field configurations which are linear in the coordinates. The latter class admits a specific subclass of isotropic solutions, where a global phase rotation of the background can be compensated by a suitable spatial rotation. We referred to the latter subclass as \emph{meta-fluids}, because they show a trivial shear elastic response alongside isotropy. 

An important feature of the model studied here is the presence of low-energy modes with reduced propagation properties. This fractonic behavior can be associated to enhanced polynomial shift symmetries of the low-energy, linear effective theory and translates into the trivialization of some elastic coefficients. More specifically, we have encountered both completely immobile modes and subdimensional modes, like \emph{lineons} propagating only along one spatial direction. The immobile fractons can be thought intuitively as \emph{plastic} deformations which cost zero energy, corresponding to an enlarged vacuum degeneracy, a property which can be compared to the diverging zero-temperature entropy of some fractonic lattice models in their continuum limit \cite{Seiberg_2021}.

Despite describing an elastic effective field theory with fractonic excitations, the models studied in the present paper differ from the setup where fracton-elasticity duality has been demonstrated \cite{Pretko:2018fed,pretko2019crystal}.
There, the two gapless modes of a symmetric gauge field represent the dual encoding of the transverse and longitudinal phonons.  Our models lack a gauge field and encode the phonons as Nambu-Goldstone modes, dynamically generated by the interactions. Furthermore, the fractons described in the present paper do not correspond to defects or non-perturbative configurations. They are the low-energy encoding of a trivial (or partially trivial) elastic response.

Immobile fractons cannot move when in isolation but can move due to interactions \cite{Pretko:2018fed}. Analogously, we expect that the flat fractonic dispersion relations that we encountered are in general ``bent" when considering higher non-linear terms in the effective theory, leading to non-trivial propagation.

To conclude, we briefly comment the relation of the present study with some applications in condensed matter and cosmology.

\begin{enumerate}

\item {\bfseries LOFF (or FFLO) superconductors:} The Larkin-Ovchinnikov-Fulde-Ferrell (LOFF) state \cite{larkin:1964zz,Fulde:1964zz} is characterized by a spatially modulated order parameter and a gapless phonon associated to the breaking of translation invariance, and it might be realized in some unconventional superconductors \cite{Matsuda_2007,Pfleiderer_2009}. A similar color superconducting state can also arise at high density in QCD \cite{Alford:2000ze,Casalbuoni:2001gt}. The Mexican hat model is similar to the Ginzburg-Landau functional used to describe  superconducting states in the particle-hole symmetric case \cite{BUZDIN1997341,Huang:2021bnm}. It could be interesting to revisit the description of the LOFF state and other translation-breaking superconducting states to explore possible emergent symmetries and the fractonic nature of the associated modes.%
\footnote{A holographic discussion of FFLO phases has been commented in \cite{Bigazzi:2011ak,Musso:2013rva,Musso:2013ija}.}

    \item {\bfseries Wigner crystals:}  In the low electron density regime, Wigner crystals can be treated according to classical elastic theory \cite{PhysRevB.15.1959,Littlewood1996}.
 There, the chemical potential $\mu$ can possibly be sufficiently low as to allow for a $\mu$-gapped dilaton to become relevant for the low-energy 
 collective-mode description of the crystal response. Despite the non-vanishing $\mu$, in a clean Wigner crystal, all the phonons (either longitudinal or transverse) are gapless.
 This matches with what we observed in the meta-fluid model, which however suggests that the standard elastic description for Wigner crystals could lack an extra (gapped) 
 dilatonic degree of freedom.

 \item {\bfseries Charge density waves:} Optical conductivity measurements show a rich structure of peaks \cite{PhysRevB.44.7808,Monceau_2012}. 
 In the presence of disorder, the general pattern is characterized by a low-frequency gapped mode corresponding to a pinned collective sliding mode 
 of the density wave condensate. At the opposite end of the spectrum, there is a high-frequency mode associated to the single 
 excitation through the density wave gap. The intermediate region features non-universal peaks corresponding to substrate modes,
 \emph{e.g.} due to the impurities. It would be interesting to investigate whether the intermediate structure could conceal
 a gapped dilatonic peak.%
 \footnote{An analogous question would be interesting also in relation to holographic realizations of charge density waves, see for instance \cite{Amoretti:2017frz,Amoretti:2017axe,Amoretti:2018tzw,Amoretti:2019cef,Amoretti:2019kuf}.}
 
 \item {\bfseries Ghost condensates:}
 In the search of possible infrared modifications of General Relativity, a mechanism similar to the gradient Mexican hat has been proposed in the time derivative sector, this is usually referred to as \emph{ghost condensation} \cite{Hamed_2004}.%
 \footnote{We refer to  \cite{Piazza_2004} for a discussion involving a dilatonic ghost. Ghost condensation is related to the spontaneous development of a harmonic time dependence, as such, is related to Floquet systems (see for a holographic discussion \cite{Biasi:2017kkn}).} A relativistic generalization of \eqref{lagra} is possible, yet it leads to trivial results. Specifically, one can consider the model
 \begin{equation}\label{rela}
     {\cal L} = 
     A \partial_\mu \Phi^* \partial^\mu \Phi - B \frac{(\partial_\mu \Phi^* \partial^\mu \Phi)^2}{\Xi^6}
     -\frac{1}{2}\partial_\mu \Xi \partial^\mu \Xi
     - H \Xi^6 
     - \lambda^2 (\Phi^* \Phi)^3\ .
 \end{equation}
 For $\lambda\neq 0$ the equations of motion imply $\Phi = 0$. Whereas, for $\lambda=0$, the resulting low-energy effective theory features just a relativistic gapless mode, the other two degrees of freedom in \eqref{rela} being associated to an emergent gauge redundancy at the quadratic level in the fluctuations. 
 
\end{enumerate}

\section*{Acknowledgements}
R.A.~and D.N.~would like to thank Colin Sterckx for a discussion about fractons.
R.A.~and D.N.~acknowledge support by IISN-Belgium (convention 4.4503.15) and by the F.R.S.-FNRS under the ``Excellence of Science" EOS be.h project n.~30820817. R.A.~is a Research Director of the F.R.S.-FNRS (Belgium).
C.H. has been partially supported by the Spanish {\em Ministerio de  Ciencia, Innovaci\'on y Universidades} through the grant PGC2018-096894-B-100.

\appendix 

\section{Homogeneous vacua}
\label{HomogeneousVacua}

In this appendix we motivate the choice of the two different symmetry breaking vacua that we have discussed in the main text, namely the helical superfluid and the meta-fluid.

Solutions to the equations of motion \eqref{EOM1} and \eqref{EOM2} which minimize the energy necessarily imply a constant $\Xi$ and a static $\Phi$. The space-dependence of $\Phi$ is further constrained to satisfy%
\footnote{An analogous equation is described in \cite{Alford_2013} in relation to superfluids with constant superfluid velocity.}
\begin{equation}\label{hom_con}
    \partial_i \Phi^* \partial_i \Phi = \frac{A}{2B}v^6\equiv c^2\ .
\end{equation}
In principle, to explore the space of time-independent solutions, one must consider the most general $\Phi$ satisfying \eqref{hom_con}. Since we are considering field theories with 2 spatial dimensions, the field $\Phi$ represents a map from the real plane to the complex plane. The condition \eqref{hom_con} restricts to maps whose complex gradient has constant modulus. The (functional) space of solutions is clearly very large. 

However, we will add one physically motivated constraint, which is to require that the effective theory of the fluctuations around the vacuum solution be completely homogeneous. In other words, we require the effective Lagrangian of the fluctuating fields not to have any explicit space dependent function. 

Suppose $\Phi_0(x_i)$ is a solution of \eqref{hom_con}. We expand the field around such solution as
\begin{align}
    \Phi(t,x_i)=\Phi_0(x_i)+f(x_i)\varphi(t,x_i)\ ,
\end{align}
where $\varphi$ is the fluctuating field, and $f(x_i)$ is a complex function, depending only on space coordinates, that takes into account the freedom in the definition of the fluctuating field. It will be fixed in order to have an homogeneous effective Lagrangian. 

Let us first consider the term with the time derivatives:
\begin{align}\label{time}
    \partial_t \Phi^*\partial_t\Phi=|f|^2 \partial_t \varphi^*\partial_t\varphi\ .
\end{align}
Homogeneity is achieved requiring $|f|^2$ to be spacetime independent. Hence, $f$ can be considered to have only a space-dependent phase.

Consider now the expansion of the expression squaring to the `gradient Mexican hat,' to linear order in the fluctuations:
\begin{align}
    \partial_i \Phi^* \partial_i \Phi-c^2 &=\partial_i \Phi_0^* \partial_i \Phi_0+\partial_i\Phi_0^*(\partial_if\varphi +f\partial_i\varphi)+\partial_i\Phi_0(\partial_if^*\varphi^*+f^*\partial_i\varphi^*)-c^2\nonumber\\
    &=\partial_i\Phi_0^*\partial_if\varphi+\partial_i\Phi_0\partial_if^*\varphi^*+\partial_i\Phi_0^*f\partial_i\varphi+\partial_i\Phi_0f^*\partial_i\varphi^*\ .
\end{align}
The quadratic Lagrangian involves the square of the above expression, and will be homogeneous if and only if each coefficient of the four terms above is itself space-independent (or zero). Taking into account that they come in complex pairs, we have the two conditions relating $f$ and $\Phi_0$:
\begin{align}\label{fcond}
    f\partial_i\Phi_0^*=ia_i\ ,  \qquad \partial_if\partial_i\Phi_0^*=b\ ,
\end{align}
where $a_i$ and $b$ are generic complex space-independent constants.

Let us now implement the fact that $f$ must have all its space-dependence in a real phase: 
\begin{align}
    f(x_i)=f_0 e^{i\theta(x_i)}\ .
\end{align}
From the first of \eqref{fcond} we get
\begin{align}
    \partial_i\Phi_0^*=i\frac{a_i}{f_0}e^{-i\theta}\ .
\end{align}
From the fact that $\partial_i\partial_j\Phi_0^*=\partial_j\partial_i\Phi_0^*$ we get that
\begin{align}
    a_i\partial_j\theta=a_j\partial_i\theta\ .
\end{align}
The second of \eqref{fcond} gives now
\begin{align}
    a_i\partial_i\theta = -b\ .
\end{align}
These last two sets of equations imply that $\partial_i\theta$ are both constant (and must be real for consistency). 

If at least one of the constants is not zero (i.e.~$b\neq0$), then we can write $\theta=k_ix_i$, and we have $f\propto \Phi_0 = \rho\, e^{ik_ix_i}$, i.e.~the helical solution (rotated towards a generic direction).

If on the other hand both constants are zero (i.e.~$b=0$), then $\theta$ is a constant that can be reabsorbed in $f_0$, the constant value of $f$. Then $\Phi_0$ is linear, $\Phi_0=b_ix_i$, i.e.~we have the meta-fluid solution (generalized to the non-isotropic case).

\section{Ward-Takahashi identities}\label{sec:WT}

The model in $2+1$ dimensions has a Lagrangian density
\begin{equation}
\cL=\cL(\partial_0 X^I, \partial_i X^I)\ ,\quad \ X^I=\{\,\Phi,\Phi^*,\Xi\}\ ,
\end{equation}
where
\begin{equation}
\begin{split}
&\cL=|\partial_0\Phi|^2+\frac{1}{2}(\partial_0\Xi)^2+A|\partial_k\Phi|^2-\frac{1}{2}(\partial_k\Xi)^2-H\Xi^6\\
&-B\Xi^{-6}|\partial_k\Phi|^4+G\Xi^{-6}|\partial_k\Phi^*\partial_k\Phi^*|^2.
\end{split}
\end{equation}
The Noether energy-momentum tensor is
\begin{equation}
T^\mu_{\ \nu}=\frac{\delta \cL}{\delta \partial_\mu X^I}\partial_\nu X^I-\delta^\mu_\nu \cL.
\end{equation}
One can check that it is conserved on-shell $\partial_\mu T^\mu_{\ \nu}=0$.
The spatial components are symmetric 
\begin{equation}
\begin{split}
&T_{ij}^{(0)}=2A\partial_{(i}\Phi^*\partial_{j)}\Phi-\partial_i \Xi \partial_j \Xi-\delta_{ij}\cL,\\
&T_{ij}^B =-4B \Xi^{-6}(\partial_k\Phi^*\partial_k\Phi)\partial_{(i}\Phi^*\partial_{j)}\Phi,\\
&T_{ij}^G=2G\Xi^{-6}\left[ (\partial_k\Phi^*\partial_k\Phi^*)\partial_i \Phi \partial_j\Phi+c.c.\right],
\end{split}
\end{equation}
so the complete stress tensor is
\begin{equation}
T_{ij}=T_{ij}^{(0)}+T_{ij}^B+T_{ij}^G.
\end{equation}
The $T_{00}$ component is
\begin{equation}
T_{00}=2\partial_0\Phi^*\partial_0 \Phi+(\partial_0\Xi)^2-\cL.
\end{equation}
Then, the trace is
\begin{equation}
T^\mu_{\ \mu}=T_{00}+\delta^{ij}T_{ij}.
\end{equation}
The traces are
\begin{equation}
\begin{split}
&\delta^{ij}T_{ij}^{(0)}=2A\partial_k\Phi^*\partial_k\Phi-\partial_k \Xi \partial_k \Xi-2\cL,\\
&\delta^{ij}T_{ij}^B =-4B \Xi^{-6}(\partial_k\Phi^*\partial_k\Phi)^2,\\
&\delta^{ij}T_{ij}^G=4G\Xi^{-6}|\partial_k\Phi^*\partial_k\Phi^*|^2,\\
\end{split}
\end{equation}
All together
\begin{equation}
\begin{split}
&T^\mu_{\ \mu}=-|\partial_0\Phi|^2-\frac{1}{2}(\partial_0\Xi)^2-A|\partial_k\Phi|^2+\frac{1}{2}(\partial_k\Xi)^2+3H\Xi^6\\
&-B\Xi^{-6}|\partial_k\Phi|^4+G\Xi^{-6}|\partial_k\Phi^*\partial_k\Phi^*|^2.
\end{split}
\end{equation}
Using the equation of motion for $\Xi$, we can write this as
\begin{equation}
\begin{split}
&T^\mu_{\ \mu}=-|\partial_0\Phi|^2-A|\partial_k\Phi|^2-\frac{1}{4}\left( \partial_0^2-\partial_k^2\right)\Xi^2\\
&+2B\Xi^{-6}|\partial_k\Phi|^4-2G\Xi^{-6}|\partial_k\Phi^*\partial_k\Phi^*|^2.
\end{split}
\end{equation}
Using now the equations of motion for $\Phi$, $\Phi^*$
\begin{equation}
\begin{split}
&T^\mu_{\ \mu}=-\frac{1}{2}(\partial_0^2+A\partial_k^2)|\Phi|^2-\frac{1}{4}\left( \partial_0^2-\partial_k^2\right)\Xi^2\\
&+\partial_k\left(B\Xi^{-6}|\partial_m\Phi|^2\partial_k|\Phi|^2\right)-\frac{1}{2}\partial_k\left[G\Xi^{-6}(\partial_m\Phi^*\partial_m\Phi^*)\partial_k\Phi^2+c.c.\right].
\end{split}
\end{equation}
We can partially improve the energy momentum tensor
\begin{equation}\label{eq:improv}
\cT^\mu_{\ \nu}=T^\mu_{\ \nu}+\frac{1}{4}\left(\square \delta^\mu_\nu-\partial^\mu \partial_\nu \right)\left( |\Phi|^2 +\frac{\Xi^2}{2}\right)+\theta^\mu_{\ \nu}.
\end{equation}
Where $\partial^\mu=\eta^{\mu\alpha}\partial_\alpha$, $\square=\eta^{\alpha\beta}\partial_\alpha\partial_\beta$ and the non-zero components of $\theta^\mu_{\ \nu}$ are
\begin{equation}
\theta^i_{\ j}=\frac{1}{2}(A+1)\left( \partial_k^2\delta^i_{\ j}-\partial_i \partial_j\right)|\Phi|^2.
\end{equation}
Then, the trace is
\begin{equation}\label{eq:wardid}
\cT^\mu_{\ \mu}=\partial^\mu V_\mu,
\end{equation}
where $V^0=0$ and
\begin{equation}
\begin{split}
&V_i=B\Xi^{-6}|\partial_k\Phi|^2\partial_i|\Phi|^2-\frac{1}{2}\left[G\Xi^{-6}(\partial_k\Phi^*\partial_k\Phi^*)\partial_i\Phi^2+c.c.\right].
\end{split}
\end{equation}

There is a conserved current associated to scale transformations
\begin{equation}
D^\mu=\cT^\mu_{\ \alpha }x^\alpha-V^\mu,\ \ \partial_\mu D^\mu=\cT^\mu_{\ \mu}-\partial^\mu V_\mu=0. 
\end{equation}
Then, \eqref{eq:wardid} is the Ward-Takahashi identity associated to dilatations.

\subsection{Conserved current}

The current is
\begin{equation}
J^\mu=\frac{i}{2}\left[ \Phi\frac{\delta \cL}{\delta \partial_\mu \Phi}- c.c.\right].
\end{equation}
The ordinary current is
\begin{equation}
j_\mu=\frac{i}{2}\left( \Phi \partial_\mu \Phi^*-\Phi^*\partial_\mu \Phi\right).
\end{equation}
In this model, the components of the conserved current are
\begin{equation}
\begin{split}
&J_0=j_0,\\ 
&J_i=-\left(A-2 B \Xi^{-6}|\partial_k\Phi|^2\right)j_i-iG\Xi^{-6}\left[(\partial_k\Phi^*)^2\Phi\partial_i \Phi -(\partial_k\Phi)^2\Phi^*\partial_i \Phi^* \right].
\end{split}
\end{equation}
And the current conservation equation is
\begin{equation}
\partial^\mu J_\mu =0.
\end{equation}

\subsection{Shift symmetries}

The Lagrangian has additional shift symmetry (we consider here only real shifts, to avoid overcounting)
\begin{equation}
\Phi\to \Phi+\alpha, \qquad \Phi^*\to \Phi^*+\alpha.
\end{equation}
The Noether currents associated to this symmetry is
\begin{equation}
J_s^\mu=\frac{\delta \cL}{\delta \partial_\mu \Phi}+\frac{\delta \cL}{\delta \partial_\mu \Phi^*}.
\end{equation}
If the action only depends on derivatives of $\Phi$, then the current is conserved, since it is a combination  of the equations of motion for $\Phi$, $\Phi^*$
\begin{equation}
\partial^\mu J_{s\,\mu}=0.
\end{equation}
Note that adding this equation makes the system of equations from the Ward-Takahashi identities equal to the system of equations from the Lagrangian, we have to solve for all the modes.

The components are
\begin{equation}
\begin{split}
&J_{s\, 0}=\partial_0\Phi+\partial_0\Phi^*,\\
&J_{s\,i}=-A\left(\partial_i\Phi+\partial_i\Phi^*\right)+2B\Xi^{-6}|\partial_k\Phi|^2\left(\partial_i\Phi+\partial_i\Phi^* \right)-2G\Xi^{-6}\left( \partial_k\Phi^*\partial_k\Phi^* \partial_i\Phi+ \partial_k\Phi\partial_k\Phi \partial_i\Phi^* \right).
\end{split}
\end{equation}

\subsection{Adding a chemical potential}

We introduce a chemical potential
\begin{equation}
\Phi=e^{i\mu t}\phi,\qquad \Phi^*=e^{-i\mu t}\phi^*.
\end{equation}
Then, the charge density becomes
\begin{equation}
J_0(\Phi)=j_0(\Phi)=4\mu|\phi|^2+j_0(\phi)\equiv J_0(\phi).
\end{equation}
The time-time component of the energy-momentum tensor changes to
\begin{equation}
T_{00}(\Phi)=2\mu^2|\phi|^2+2i\mu(\phi\partial_0\phi^* -\phi^*\partial_0\phi)+T_{00}(\phi)=\mu J_0(\phi)+t_{00}(\phi),
\end{equation}
where
\begin{equation}
t_{00}(\phi)=T_{00}(\phi)-2\mu^2|\phi|^2.
\end{equation}
The effective Lagrangian is
\begin{equation}
\cL_\phi=\cL+2\mu^2|\phi|^2.
\end{equation}

The change in the trace is
\begin{equation}
T^\mu_{\ \mu}(\Phi)=T^\mu_{\ \mu}(\phi)-\mu^2|\phi|^2-i\mu (\phi\partial_0\phi^* -\phi^*\partial_0\phi).
\end{equation}

\section{Generalizations to 3+1 dimensions}

In this appendix, we briefly outline generalizations to $3+1$ dimensional systems, to show that the essential features of both the helical superfluid and the meta-fluid are unchanged. The only difference is that we have to use different models to generalize the helical superfluid and the meta-fluid, respectively. We will keep the analysis of both models to a minimum, since it turns out that they are very similar to their $2+1$ dimensional cousins.

\subsection{3+1 dimensional helical superfluid}
In order to generalize the helical superfluid, we keep the field content to be a complex scalar $\Phi$ and a real scalar $\Xi$. Only the scaling dimensions of the scalars changes, and hence the compensating powers of $\Xi$.

We thus start with the following Lagrangian, where we have already implemented a condition like \eqref{eom12}:
\begin{align}
   {\cal L} =
 \partial_t \Phi^* \partial_t \Phi
 +\frac{1}{2} \partial_t \Xi \partial_t \Xi
 -\frac{1}{2} \partial_i \Xi \partial_i \Xi
 -\frac{B}{\Xi^4}\left(\partial_i \Phi^* \partial_i \Phi-\frac{A}{2B}\Xi^4\right)^2
 \ .
\end{align}
The equations of motion are solved for
\begin{align}
    \Phi=\rho\ e^{ikx}\ , \qquad \Xi=v\ , \qquad \mbox{with} \qquad \frac{k^2\rho^2}{v^4}=\frac{A}{2B}\ .
\end{align}
The expansion is exactly as before
\begin{align}
    \Phi= \rho\ e^{ikx}(1+\sigma+i\chi)\ , \qquad \Xi = v(1+\tau)\ ,
\end{align}
so that the effective quadratic Lagrangian for the fluctuations about the helical vacuum is
\begin{align}
     {\cal L}=\frac{v^2}{2}\partial_\mu\tau\partial^\mu \tau+\rho^2(\partial_t \chi)^2+\rho^2(\partial_t\sigma)^2-2A\rho^2\left[\partial_x \chi +k(\sigma-2 \tau) \right]^2\ ,
\end{align}
which is exactly similar to \eqref{lquadheli} except for a numerical coefficient. The spectrum will then be exactly the same: there is an immobile fracton, a gapless mode which has linear and isotropic dispersion relations at low momentum, but becomes a lineon at high momentum (propagating along $x$, now one out of three directions), and a gapped mode which has relativistic dispersion relations at high momentum (this is the spectrum for $A\leq 1/2$; if $A>1/2$ then as before the lineon and the relativistic mode switch roles according to the direction of propagation). The gap is given by
\begin{align}
    m^2= 2Ak^2\left(1+8\frac{\rho^2}{v^2}\right)\ .
\end{align}

\subsection{3+1 dimensional meta-fluid}
In order to generalize the meta-fluid, the model has to contain as many (real) scalar fields as space directions, plus the compensator scalar field. Hence we start with 3 real scalar fields $\Phi_i$, to which we add $\Xi$. The Lagrangian is now
\begin{align}
   {\cal L} =
 \frac12\partial_t \Phi_i \partial_t \Phi_i
 +\frac{1}{2} \partial_t \Xi \partial_t \Xi
 -\frac{1}{2} \partial_i \Xi \partial_i \Xi
 -\frac{B}{\Xi^4}\left(\partial_i \Phi_j \partial_i \Phi_j-\frac{A}{2B}\Xi^4\right)^2
 \ .
\end{align}
The solution to the equations of motion is
\begin{align}
    \Phi_i=b\ x_i\ , \qquad \Xi=v\ , \qquad \mbox{with} \qquad \frac{3b^2}{v^4}=\frac{A}{2B}\ .
\end{align}
We take the fluctuations to be
\begin{align}
    \Phi_i=b(x_i+u_i)\ , \qquad \Xi=v+\tau\ ,
\end{align}
and the quadratic Lagrangian becomes
\begin{align}
    {\cal L}=\frac{1}{2}\partial_\mu\tau \partial^\mu\tau +\frac12b^2\partial_t u_i \partial_t u_i-\frac23Ab^2\left(\partial_i u_i-\frac6v \tau \right)^2\ .
\end{align}
Again, this is very similar to \eqref{eq:elastic1}, up to some numerical coefficients. However, now it involves 4 modes instead of three. But we can immediately see that the only modes that will have non-trivial dispersion relations are the mixtures of $\tau$ and the longitudinal component of $u_i$. Then, both transverse modes of $u_i$ will be immobile fractons. 

As for the non-trivial modes, one is gapped with gap given by 
\begin{align}
    m^2=48 A\frac{b^2}{v^2}\ ,
\end{align}
and relativistic dispersion relation at high momentum, while the other is gapless with quadratic dispersion relation at low momentum
\begin{align}
    \omega^2=\frac43\frac{A}{m^2} q^4+{\cal O}(q^6)\ ,
\end{align}
and linear dispersion relation given by
\begin{align}
    \omega^2\simeq \frac{4}{3} Aq^2 \ ,
\end{align}
at high momentum. If $A>3/4$, the high momentum behavior is switched between the two modes.

To summarize, we see that the generalization to 3+1 dimensions yields physics very similar to the 2+1 dimensional case that we have analyzed in detail, so that we expect that the latter transposes to 3+1 dimensions straightforwardly.

\bibliography{skeleton} 
\bibliographystyle{utphys}

\end{document}